\documentclass[aps, reprint, superscriptaddress, floatfix, showpacs, amsmath, amssymb]{revtex4-1}

\usepackage{graphicx}
\usepackage{dcolumn}
\usepackage{bm}
\usepackage{hyperref}

\begin{document}

\title{Longitudinal Double-Spin Asymmetries for Dijet Production \\
at Intermediate Pseudorapidity in Polarized $pp$ Collisions at $\sqrt{s}$ = 200 GeV}

\affiliation{AGH University of Science and Technology, FPACS, Cracow 30-059, Poland}
\affiliation{Argonne National Laboratory, Argonne, Illinois 60439}
\affiliation{Brookhaven National Laboratory, Upton, New York 11973}
\affiliation{University of California, Berkeley, California 94720}
\affiliation{University of California, Davis, California 95616}
\affiliation{University of California, Los Angeles, California 90095}
\affiliation{Central China Normal University, Wuhan, Hubei 430079}
\affiliation{University of Illinois at Chicago, Chicago, Illinois 60607}
\affiliation{Creighton University, Omaha, Nebraska 68178}
\affiliation{Czech Technical University in Prague, FNSPE, Prague, 115 19, Czech Republic}
\affiliation{Nuclear Physics Institute AS CR, Prague 250 68, Czech Republic}
\affiliation{Technische Universitat Darmstadt, Germany}
\affiliation{Frankfurt Institute for Advanced Studies FIAS, Frankfurt 60438, Germany}
\affiliation{Fudan University, Shanghai, 200433 China}
\affiliation{Institute of Physics, Bhubaneswar 751005, India}
\affiliation{Indiana University, Bloomington, Indiana 47408}
\affiliation{Alikhanov Institute for Theoretical and Experimental Physics, Moscow 117218, Russia}
\affiliation{University of Jammu, Jammu 180001, India}
\affiliation{Joint Institute for Nuclear Research, Dubna, 141 980, Russia}
\affiliation{Kent State University, Kent, Ohio 44242}
\affiliation{University of Kentucky, Lexington, Kentucky 40506-0055}
\affiliation{Lamar University, Physics Department, Beaumont, Texas 77710}
\affiliation{Institute of Modern Physics, Chinese Academy of Sciences, Lanzhou, Gansu 730000}
\affiliation{Lawrence Berkeley National Laboratory, Berkeley, California 94720}
\affiliation{Lehigh University, Bethlehem, Pennsylvania 18015}
\affiliation{Max-Planck-Institut fur Physik, Munich 80805, Germany}
\affiliation{Michigan State University, East Lansing, Michigan 48824}
\affiliation{National Research Nuclear University MEPhI, Moscow 115409, Russia}
\affiliation{National Institute of Science Education and Research, HBNI, Jatni 752050, India}
\affiliation{National Cheng Kung University, Tainan 70101 }
\affiliation{Ohio State University, Columbus, Ohio 43210}
\affiliation{Institute of Nuclear Physics PAN, Cracow 31-342, Poland}
\affiliation{Panjab University, Chandigarh 160014, India}
\affiliation{Pennsylvania State University, University Park, Pennsylvania 16802}
\affiliation{Institute of High Energy Physics, Protvino 142281, Russia}
\affiliation{Purdue University, West Lafayette, Indiana 47907}
\affiliation{Pusan National University, Pusan 46241, Korea}
\affiliation{Rice University, Houston, Texas 77251}
\affiliation{Rutgers University, Piscataway, New Jersey 08854}
\affiliation{Universidade de Sao Paulo, Sao Paulo, Brazil, 05314-970}
\affiliation{University of Science and Technology of China, Hefei, Anhui 230026}
\affiliation{Shandong University, Jinan, Shandong 250100}
\affiliation{Shanghai Institute of Applied Physics, Chinese Academy of Sciences, Shanghai 201800}
\affiliation{State University of New York, Stony Brook, New York 11794}
\affiliation{Temple University, Philadelphia, Pennsylvania 19122}
\affiliation{Texas A\&M University, College Station, Texas 77843}
\affiliation{University of Texas, Austin, Texas 78712}
\affiliation{University of Houston, Houston, Texas 77204}
\affiliation{Tsinghua University, Beijing 100084}
\affiliation{University of Tsukuba, Tsukuba, Ibaraki 305-8571, Japan}
\affiliation{Southern Connecticut State University, New Haven, Connecticut 06515}
\affiliation{University of California, Riverside, California 92521}
\affiliation{University of Heidelberg, Heidelberg, 69120, Germany }
\affiliation{United States Naval Academy, Annapolis, Maryland 21402}
\affiliation{Valparaiso University, Valparaiso, Indiana 46383}
\affiliation{Variable Energy Cyclotron Centre, Kolkata 700064, India}
\affiliation{Warsaw University of Technology, Warsaw 00-661, Poland}
\affiliation{Wayne State University, Detroit, Michigan 48201}
\affiliation{Yale University, New Haven, Connecticut 06520}

\author{J.~Adam}\affiliation{Creighton University, Omaha, Nebraska 68178}
\author{L.~Adamczyk}\affiliation{AGH University of Science and Technology, FPACS, Cracow 30-059, Poland}
\author{J.~R.~Adams}\affiliation{Ohio State University, Columbus, Ohio 43210}
\author{J.~K.~Adkins}\affiliation{University of Kentucky, Lexington, Kentucky 40506-0055}
\author{G.~Agakishiev}\affiliation{Joint Institute for Nuclear Research, Dubna, 141 980, Russia}
\author{M.~M.~Aggarwal}\affiliation{Panjab University, Chandigarh 160014, India}
\author{Z.~Ahammed}\affiliation{Variable Energy Cyclotron Centre, Kolkata 700064, India}
\author{N.~N.~Ajitanand}\affiliation{State University of New York, Stony Brook, New York 11794}
\author{I.~Alekseev}\affiliation{Alikhanov Institute for Theoretical and Experimental Physics, Moscow 117218, Russia}\affiliation{National Research Nuclear University MEPhI, Moscow 115409, Russia}
\author{D.~M.~Anderson}\affiliation{Texas A\&M University, College Station, Texas 77843}
% ...

\collaboration{STAR Collaboration}\noaffiliation

\date{\today}

\begin{abstract}
We present the first measurements of the longitudinal double-spin asymmetry $A_{LL}$ for dijets with at least one jet reconstructed within the pseudorapidity range $0.8 < \eta < 1.8$. The dijets were measured in polarized $pp$ collisions at a center-of-mass energy $\sqrt{s}$ = 200 GeV. Values for $A_{LL}$ are determined for several distinct event topologies, defined by the jet pseudorapidities, and span a range of parton momentum fraction $x$ down to $x \sim$ 0.01. The measured asymmetries are found to be consistent with the predictions of global analyses that incorporate the results of previous RHIC measurements. They will provide new constraints on $\Delta g(x)$ in this poorly constrained region when included in future global analyses.
\end{abstract}

\pacs{13.87.Ce, 13.88.+e, 14.20.Dh, 14.70.Dj}

\maketitle

\section{Introduction and Motivation}

Understanding the internal spin structure of the proton is a fundamental goal in strong interaction physics. Deep inelastic lepton scattering (DIS) measurements have played a seminal role in the development of our present knowledge of hadronic substructure. Studies of {\em polarized\/} deep inelastic lepton scattering (pDIS), in which a longitudinally-polarized lepton beam scatters from a longitudinally or transversely polarized target, have provided important insights into the spin structure of the nucleon. Several decades of increasingly precise pDIS experiments have found that the spins of the quarks $(\Delta \Sigma)$ account for only $\sim30\%$ of the total spin of the proton, with the remainder due to contributions from the gluon spin $(\Delta G)$ and the orbital angular momenta ($L$) of the partons (\cite{Nocera:2014gqa, deFlorian:2014yva} and references therein).\par

The helicity distribution of gluons within the proton, $\Delta g(x)$, is thus a key ingredient in unraveling the internal structure and the QCD dynamics of nucleons. The Relativistic Heavy Ion Collider (RHIC) \cite{Alekseev:2003sk} at Brookhaven National Laboratory is a unique tool for exploring gluon polarization, through collisions of polarized proton beams at center-of-mass energies $\sqrt{s}=200$ and 510 GeV. At these energies, RHIC kinematics is particularly sensitive to gluons, as scattering occurs predominantly via quark-gluon and gluon-gluon interactions.\par

Previous measurements of the longitudinal double-spin asymmetries, $A_{LL}$, for inclusive jet \cite{Abelev:2006uq, Abelev:2007vt, Adamczyk:2012qj, Adamczyk:2015} and $\pi^{0}$ \cite{Adare:2008aa, Adare:2008qb, Adare:2014hsq} production, obtained by the STAR and PHENIX experiments at RHIC respectively, have been added to the DSSV \cite{deFlorian:2014yva} global analyses. Inclusive jets \cite{Abelev:2006uq, Abelev:2007vt, Adamczyk:2012qj, Adamczyk:2015, Adare:2010cc} measurements were included in NNPDF \cite{Nocera:2014gqa} global analyses. The addition of the most recent STAR inclusive jet results \cite{Adamczyk:2015} shows, for the first time, a positive gluon polarization in the region of sensitivity, $x > 0.05$. At lower values of the momentum fraction $x$, however, the magnitude and shape of the gluon helicity distribution are still poorly constrained. \par

Correlation observables, such as those from dijet production, capture more information about the initial state kinematics of the hard scattering, and may lead to tighter constraints on the shape of $\Delta g(x)$. Recently, STAR published the cross section and first measurements of $A_{LL}$ for dijets produced near mid-rapidity in longitudinally-polarized proton-proton collisions at $\sqrt{s}=200$ GeV \cite{Adamczyk:2016okk}. The measured cross section was found to be consistent with next-to-leading order (NLO) perturbative QCD expectations. The extracted spin asymmetries also showed good agreement with the predictions of current global analyses \cite{Nocera:2014gqa, deFlorian:2014yva}. The dijet invariant mass is proportional to the square-root of the product of the initial state momentum fractions, $M = \sqrt{s}\sqrt{x_{1}x_{2}}$, at leading order QCD; and the sum of the jet pseudorapidities determines their ratio, $\eta_{3}+\eta_{4}=\ln{(x_{1}/x_{2})}$, where we follow the convention that the initial (final) state kinematics are referenced with index 1,2 (3,4). Adding dijet results to the global analyses will further constrain the $x$ dependence of $\Delta g$.\par

In this paper, we report the first measurements of the longitudinal double-spin asymmetry, $A_{LL}$, for dijet production at {\em intermediate\/} pseudorapidities, where at least one of the jets was detected in the range of $0.8 < \eta < 1.8$. The data were taken at $\sqrt{s}=200$ GeV in 2009 by the STAR collaboration, and extend the sensitivity to parton distributions at lower $x$ values than those probed at mid-rapidity \cite{Adamczyk:2016okk}.\par

The remainder of this paper is organized as follows: in Sec.\ II we briefly describe relevant aspects of the STAR detector; Sec. III discusses the data and simulation samples used; Sec.\ IV focuses on our jet reconstruction and selection criteria, while Sec.\ V provides details on the experimental methods. The double spin asymmetry $A_{LL}$ measurements are presented in Sec.\ VI, and the associated bias and uncertainties are discussed in Sec.\ VII. The results are presented in Sec.\ VIII, with our summary in Sec.\ IX.

\section{The STAR Detector at RHIC}
RHIC consists of two quasi-circular concentric accelerator/storage rings on a common horizontal plane, one (`Blue Ring') for clockwise and the other (`Yellow Ring') for counter-clockwise beams. Each ring can store 120 proton bunches. The overall efficiency of the acceleration process and beam transfer into RHIC is higher than $50\%$, yielding  about $2 \times 10^{11}$ protons per bunch. The (vertical) polarizations of the proton beams are maintained by use of `Siberian Snakes', and are measured several times per fill, as discussed in Sec.\ III.A and VI.A. Spin rotator magnets, located on each side of the two major interaction points, can precess the stable spin orientation from vertical into the horizontal plane, and back, allowing for collisions of longitudinally polarized beams \cite{Alekseev:2003sk}.\par

The Solenoidal Tracker at RHIC (STAR) \cite{Ackermann:2002ad} is a multipurpose detector designed to measure hadronic and electromagnetic particles in heavy-ion and polarized proton-proton collisions. STAR is comprised several subsystems which provide charged particle tracking and electromagnetic calorimetry over a wide range of pseudorapidity. The three primary subsystems used for jet reconstruction in this work are the time projection chamber (TPC) \cite{Anderson:2003ur}, the barrel electromagnetic calorimeter (BEMC) \cite{Beddo:2002zx}, and the endcap electromagnetic calorimeter (EEMC) \cite{Allgower:2002zy}. Additionally, the beam-beam counters (BBC) \cite{Kiryluk:2005gg} and zero-degree calorimeters (ZDC) \cite{Adler:2000bd} were used to determine the relative integrated luminosities of the various beam-spin combinations.\par 

The TPC provides charged-particle tracking in a 0.5~T solenoidal magnetic field over the nominal range $|\eta| \le 1.3$ in pseudorapidity and $2\pi$ in azimuthal angle. The TPC is used to determine the transverse momentum, $p_{T}$, of the outgoing charged particles, and also aids in locating the position of the collision vertex. The tracking efficiency is $\sim85\%$ for $|\eta| \le 1.0$, but falls to $\sim50\%$ at $|\eta| \sim 1.3$ \cite{Anderson:2003ur}. This is a critical issue when attempting to reconstruct jets at intermediate pseudorapidities.\par                                                                                             
Surrounding the TPC in azimuth, for the range $|\eta| < 1$, is the BEMC \cite{Beddo:2002zx}, which measures electromagnetic energy deposition. The BEMC is a lead-scintillator sampling calorimeter which is roughly 20 radiation lengths deep and consists of 4800 optically isolated projective towers, each subtending 0.05 radians in azimuth and 0.05 units in pseudorapidity.\par

The EEMC \cite{Allgower:2002zy} is located on the west end of the TPC, and extends the kinematic reach of the BEMC in the forward direction. Like its counterpart, the EEMC is a lead-scintillator sampling calorimeter, and provides electromagnetic calorimetry for $1.09 < \eta < 2.00$ and over the full range in azimuth (there is a small service gap between the two detectors for $1.00 < \eta < 1.08$). In addition to calorimetry, both the BEMC and EEMC are used to generate the primary jet trigger information at STAR, as described in the next section.\par

\section{Data and Simulation Samples}

\subsection{Data Sets and Triggering}
The data used in this analysis were collected by the STAR collaboration in 2009, from longitudinally polarized $pp$ collisions at $\sqrt{s}$ = 200~GeV. The data set has an integrated luminosity of 21 pb$^{-1}$. Values of the proton beam polarization were extracted from the spin-dependent asymmetries observed in proton elastic scattering in the Coulomb-Nuclear Interference (CNI) region, with high-statistics measurements carried out using proton-Carbon (pC) polarimeters~\cite{Jinnouchi:2004up}, which were normalized with respect to a polarized hydrogen gas jet (H-Jet) polarimeter~\cite{Zelenski:2005mz}. The luminosity-weighted polarizations of the two beams were $P_{B} = 56\%$ and $P_{Y} = 57\%$. The relative uncertainty of the product $P_{B}P_{Y}$, relevant for this analysis, was $6.5\%$. Ratios of the integrated luminosities for different beam helicity states were determined by the BBCs \cite{Kiryluk:2005gg} and the ZDCs \cite{Adler:2000bd}. Details on these quantities, and their estimated uncertainty contributions to $A_{LL}$, are discussed in Sec.\ VI.\par

Events used in this analysis needed to pass at least one of several trigger conditions. The STAR trigger system \cite{Bieser:2002ah}, designed to optimize both the heavy-ion and spin physics programs, is a multi-level, modular, pipelined system in which digitized signals from the fast trigger detectors are examined at the RHIC crossing rate of $\sim$9~MHz. This low-level information is then used to determine whether to read out data from the slower, more finely-grained detectors and transfer all data to disk, or to reset and wait for the next event.\par

The triggers for the selection of jet events were constructed by requiring substantial energy to be present in the BEMC or EEMC within fixed $\Delta \eta \times \Delta \phi = 1\times1$ calorimeter regions (jet patches). There are a total of 18 non-overlapping jet patches that cover the BEMC and EEMC: six each in the East and West halves of the BEMC, and the remaining six in the EEMC. Since these jet patches are fixed in the detector, and are comparable in area to that of a typical jet, there are sizable triggering inefficiencies at the jet-patch boundaries. A jet that strikes near the boundary of two jet patches and shares its energy between them, for example, may not deposit enough energy in either jet patch to exceed the trigger threshold. To mitigate this effect in the $\eta$ direction, two sets of six `overlap' jet patches were created. One set straddles the boundary between the jet patches that cover a given $\phi$ range in the East and West halves of the BEMC, which meet at $\eta = 0$, while the other set straddles the boundary between the jet patches in the West half of the BEMC and those in the EEMC, which meet at $\eta \sim 1$. \par

Including the 12 overlap jet patches yielded a total of 30 jet patches available for triggering in the 2009 run configuration. Hardware restrictions prevented the implementation of analogous overlapping jet patches in the $\phi$ direction, but the inefficiencies in $\phi$ are eased by the Adjacent Jet Patch (AJP) logic. For the 2009 run, each jet patch had three associated energy thresholds: a jet patch trigger was satisfied if the transverse energy detected in a single jet patch exceeded either 5.4 GeV (the JP1 trigger, which was prescaled) or 7.3 GeV (JP2 trigger), or if two jet patches adjacent in azimuth each exceeded 3.5 GeV (the AJP trigger). The AJP logic was not implemented for the jet patches which span the service gap between the BEMC and EEMC.\par

\subsection{Simulation Samples}
Simulated events are needed to correct for detector effects on the measured quantities of interest, as well as to evaluate various systematic uncertainties. These events were generated using \textsc{Pythia} 6.4.26 \cite{Sjostrand:2006za} with the Perugia 0 tune \cite{Skands:2010ak} and were then processed through a STAR detector response package implemented in GEANT 3 \cite{geant}. The simulated events were embedded into ‘zero-bias’ events from real data, which come from triggering on random bunch crossings over the span of the run. The real and simulated events were combined at the `raw' detector level, {\it i.e.}, before the TPC padrow data is converted into track segments. This way the simulation sample can accurately mimic the same beam background, pile-up, and detector conditions as the real data throughout the entire data collection period.\par

A significant amount of computing time is needed to fully simulate and reconstruct the STAR detector response to each event generated in \textsc{Pythia}. In order to reduce the time required to run the simulation, a trigger filter was applied. The trigger filter rejects events which would not have fired the JP1 or AJP trigger. For the 2009 simulation sample, the trigger filter rejected about $91.5\%$ of all \textsc{Pythia} events; however, the full \textsc{Pythia} record for the rejected events was saved, so that corrections to the unbiased sample may be made, which will be discussed later.\par

The simulation provides three distinct levels of information. These are the partonic hard scattering, the final-state particles from the hadronization of the partons, and the response of the detector to those particles. These divisions will be referred to as the parton level, particle level, and detector level information, respectively. The parton level of the simulation contains information about the partons involved in the $2\rightarrow 2$ hard scattering event generated by \textsc{Pythia}. Various kinematic properties of the hard scattering, such as the $Q^{2}$, center-of-mass scattering angle, and momentum fractions $x$ of the incoming partons are stored. For jets reconstructed at the parton level, only the partons involved in the hard scattering and partons which arise from initial or final state radiation are used as input to the jet finding algorithm. Partons due to the underlying event or beam remnants, which arise from soft processes involving partons in the colliding protons other than the hard-scattered pair, are not included in the parton-level jet finding.\par

The partons generated by \textsc{Pythia} propagate and hadronize to form stable, color-neutral particles. The particle level of the simulation records the kinematic information and particle identification. Particle level jets are constructed using all stable particles, including those which arise from the underlying event and beam remnants.\par

The last level of the simulation records the raw response of the individual detector subsystems to the stable particles formed at the previous level. As the particles traverse the GEANT model of the detector, they interact in the various volumes consistent with the interaction of the particular particle in a specific material. This interaction includes processes such as ionizing the gas in the TPC and depositing energy in the scintillator layers of the calorimeters. This, along with a detailed simulation of the detector readout electronics and trigger logic, allows the simulation routines to generate event data which are consistent with that of the real detector. When the jet finder is run on the detector level simulation, it constructs jets from the simulated response of the TPC and calorimeter towers, as would be recorded by  their readout electronics.\par

\section{Jet Reconstruction and Event Selection}

\subsection{Jet reconstruction}
The jet reconstruction procedures used here generally follow those of the inclusive jet \cite{Adamczyk:2015} and mid-rapidity dijet \cite{Adamczyk:2016okk} analyses of the 2009 data. Jets were reconstructed using the anti-$k_{T}$ algorithm \cite{Cacciari:2008gp} implemented in the FastJet package \cite{Cacciari:2011ma} with resolution parameter $R$ = 0.6. Information input to the jet finder included charged tracks from the TPC and calorimeter tower energy deposits. Tracks were required to have $p_{T} \ge 0.2$ GeV/$c$, and individual calorimeter towers needed an $E_{T}$ which exceeded 0.2 GeV. Valid charged tracks were also required to contain more than five fit points in the TPC (see below) and at least $51\%$ of the maximum number of fit points allowed by the TPC geometry and active electronic channels. Finally, tracks were required to satisfy a $p_T$-dependent condition on the distance of closest approach (DCA), which is the minimum distance between the event vertex and any point along the track trajectory. Tracks with $p_T$ below 0.5 GeV/$c$ were required to have a DCA $<$ 2~cm, while tracks with $p_T$ above 1.5 GeV/$c$ were required to have a DCA $<$ 1~cm, with a linear interpolation between these two distances in the intermediate $p_{T}$ region. The DCA cut serves to remove pile-up tracks not associated with the hard scattering event.\par

The tracks were reconstructed from ionization along the path of a particle in the TPC volume. Electrons from the ionization drift towards the readout pads where they create a charge avalanche. These pads are situated in rows (padrows) oriented roughly perpendicular to a straight radial line emanating from the interaction point. A ``fit point" is a padrow that contributes to a reconstructed track. The condition used in this analysis on the number of fit points differs from the 2009 inclusive jet analysis, which required that tracks have more than 12 hits in order to be reconstructed. Tracks with pseudorapidity $\eta > 1$ would not traverse the entire radial extent of the endcap before leaving the TPC, so the outermost padrows will not collect any charge, leading to a smaller number of possible fit points at high pseudorapidity. Reducing the number of required hits allows more tracks to be included in the jet reconstruction. The lower five-point tracking requirement does not extend over the full TPC, and is only implemented for tracks with $\eta > 0.6$.\par

For input into the jet-finder, charged particle tracks and calorimeter tower energy deposits are converted into Lorentz invariant four-momentum vectors. The tracks are assumed to be charged pions and are assigned the pion mass, while the particles detected in the calorimeter towers are assumed to be massless. To avoid double-counting energy contributions from the TPC and the calorimeters, all towers that had tracks passing through them had the $p_{T}$ of the track subtracted from the $E_{T}$ of the tower. If the track $p_T$ was greater than the transverse energy of the tower, the tower $E_T$ was set to zero. This method has been shown to reduce the residual jet momentum corrections and the sensitivity to fluctuations in the hadronic energy deposition, resulting in an improved jet momentum resolution \cite{Adamczyk:2015}.\par

\subsection{Dijet Selection}
\begin{figure*}[!hbt]
    \includegraphics[width=2.0\columnwidth]{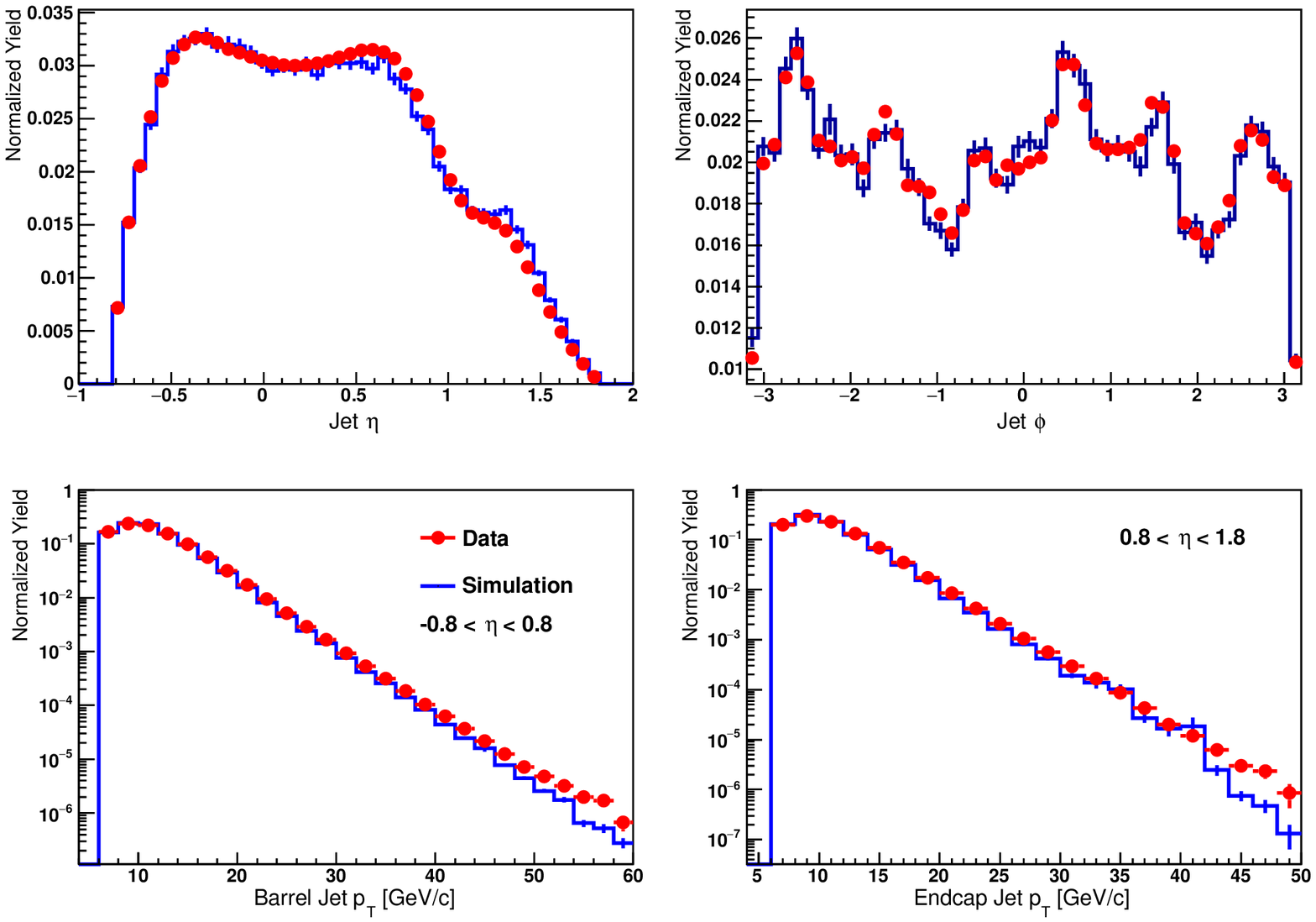}
    \caption{Data/simulation comparisons of the relative jet yields as functions of Barrel+Endcap jet pseudorapidity (upper left) and jet azimuthal angle (upper right), or as functions of detector jet transverse momentum for the Barrel (lower left) and Endcap (lower right). The solid circle points represent the data, and the histograms are the simulation.}
    \label{fig:jet_PtEtaPhi}
\end{figure*}

\begin{figure}[!hbt]
    \includegraphics[width=1.0\columnwidth]{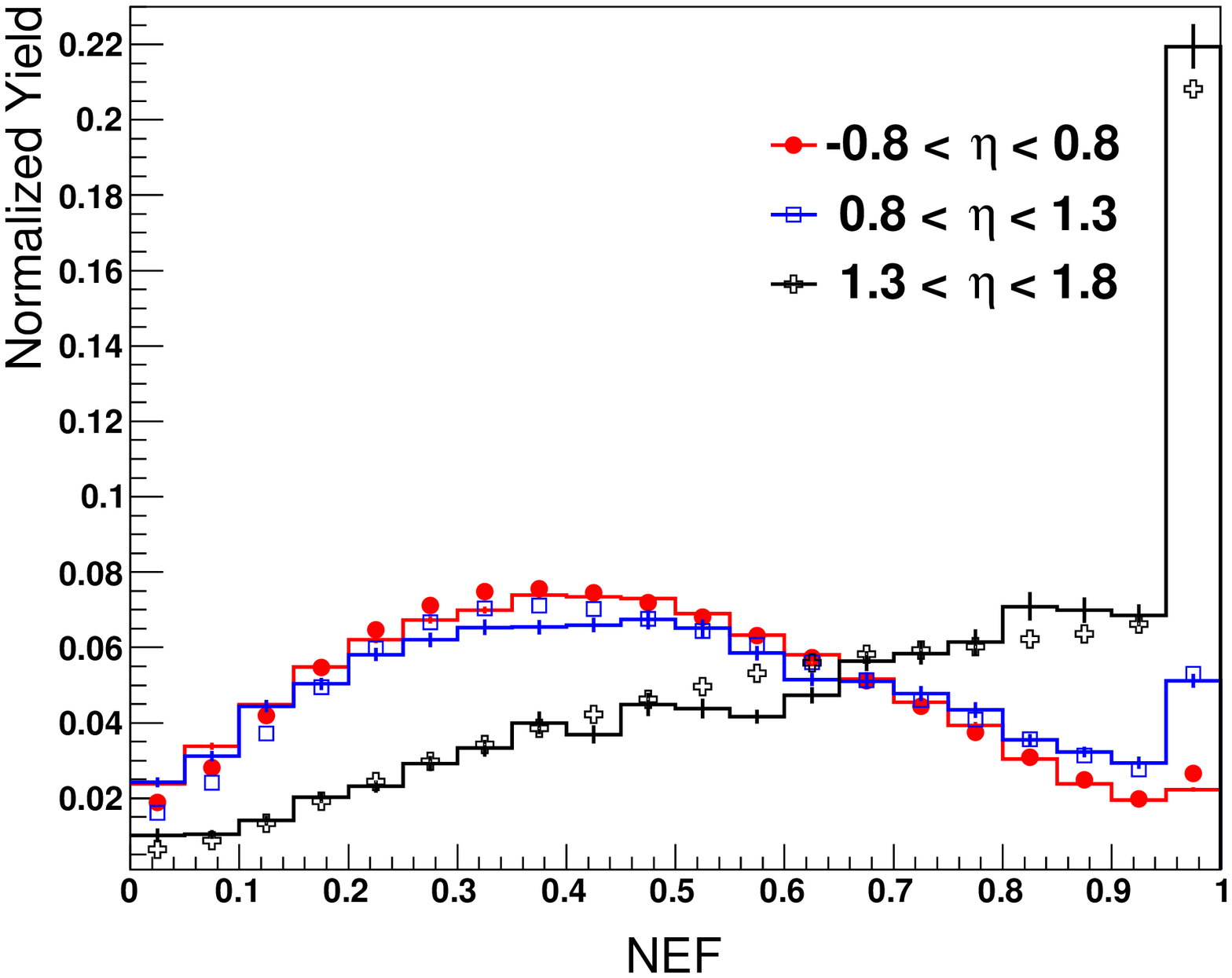}
    \caption{Data/simulation comparisons of the jet yields vs. jet neutral energy fraction (NEF), shown separately for jets in different pseudorapidity ranges. The points represent the data (solid circle for Barrel, open square and open cross for Endcap jets at two different pseudorapidity ranges), and the histogram is the simulation.}
    \label{fig:jet_Rt}
\end{figure}
The dijet selection procedure follows closely that used in the STAR 2009 mid-rapidity dijet measurement \cite{Adamczyk:2016okk}. For each event that has a $z$ vertex position within 90 cm of the center of the STAR detector, a dijet was selected by choosing the two jets with the highest $p_{T}$ that fell in the pseudorapidity range $-0.8 \le \eta \le 1.8$ and detector pseudorapidity range $-0.7 \le \eta_{\rm Det} \le 1.7$. The detector pseudorapidity is defined by extrapolating the jet thrust axis into the BEMC or EEMC detector, then calculating the pseudorapidity of that intersection point relative to the center ($z=0$) of the STAR detector. In the discussion that follows, jets with pseudorapidities  $-0.8 \le \eta \le 0.8$ will be referred as ``Barrel jets", while those in the range $0.8 \le \eta \le 1.8$ will be denoted as ``Endcap jets."\par 

The two jets arising from a partonic hard-scattering event should be roughly back-to-back in azimuth ($\phi$). Jets which are too close to each other in azimuth likely do not originate from a $2 \rightarrow 2$ hard scattering process. To remove these events from the analysis, an opening angle cut was placed on the two jets of the dijet event, such that the azimuthal angle between them must be more than $120^{\circ}$.\par

To facilitate comparison with theoretical predictions, an asymmetric condition was placed on the transverse momentum of the jets, requiring a transverse momentum of $p_T \ge$ 8.0 GeV/$c$ for one jet and $p_T \ge $ 6.0 GeV/$c$ for the other in the dijet pair~\cite{Frixione:1997ks}. Also, events containing a track with $p_T$ above 30 GeV/$c$ were removed if the jets comprising the dijet had highly imbalanced transverse momenta ($p_T$ ratios greater than 3/2 or less than 2/3). These highly imbalanced events are likely due to the finite resolution in the track curvature calculation, which will occasionally result in a significant overestimate of a track $p_T$. It was also required that at least one jet falls within the acceptance of a jet patch that satisfied the JP2, JP1, or AJP trigger.\par

In the inclusive jet analyses at STAR, a cut on the neutral energy fraction (NEF) of the jets was imposed in order to remove jets comprised primarily of background particles, due predominantly to interactions of the beam(s) with RHIC ring elements far upstream. The cut was usually placed such that jets with more than $95\%$ of their transverse momentum coming from the calorimeter towers were rejected. This requirement can not be applied when studying jets at forward pseudorapidity, as the falling TPC efficiency in this region means that the reconstructed jets will have increasingly fewer tracks, and therefore large neutral fractions. It is highly unlikely, though, that a `background jet' will be coincident with a physics jet. So rather than placing a neutral energy cut on the individual jets, the requirement was loosened to only reject dijet candidates for which both jets had neutral fractions of 1.\par

\subsection{Comparison to Simulation}

\begin{figure*}[!ht]
  \centering
    \includegraphics[width=2.0\columnwidth]{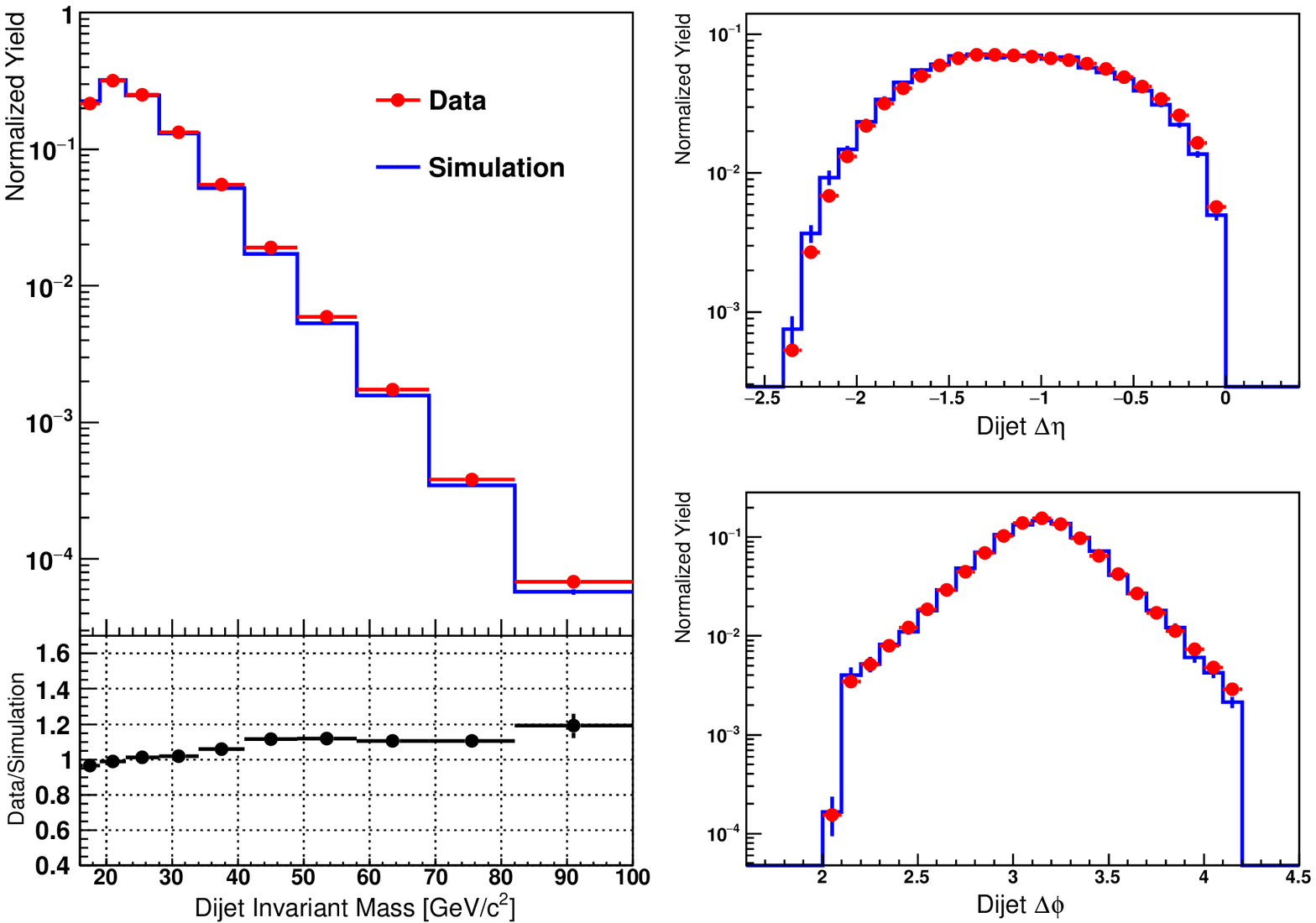}
    \caption{Data/simulation comparisons of dijet yields as a function of invariant mass (left) for all the accepted events. Similar comparisons for Barrel-Endcap dijets only as function of the pseudorapidity gap (upper right) and azimuthal opening angle (lower right) between the jets. The points are the data and the histogram is the simulation.}
    \label{fig:Dijet_Mass_DataSimu}
\end{figure*}

For the simulation sample, detector-level dijets were reconstructed from the simulated TPC and calorimeter responses using the same algorithms as were used for the data. The upper two panels in Fig.~\ref{fig:jet_PtEtaPhi} show the comparisons between data and simulation for the jet pseudorapidity and jet azimuthal angle distributions. The good agreement seen between data and simulation for jet $\eta$ and $\phi$ shows that the detector conditions are well reproduced in the simulation, as the $\phi$ spectrum in particular is sensitive to the trigger granularity and hardware readout failures in the TPC. The lower two panels show comparisons of data and simulation for jet $p_T$ spectra, separated between Barrel jets ($|\eta| <$ 0.8) and Endcap jets (0.8 $< \eta <$ 1.8). Figure~\ref{fig:jet_Rt} compares data and simulation for the observed neutral energy fraction distributions, again separately for jets in the Barrel and Endcap electromagnetic calorimeters at different pseudorapidity ranges. They show good agreement. The shift to higher neutral fraction, {\it i.e.}, to a larger fraction of the jet energy detected in the calorimeters, is apparent for the Endcap, reflecting the decreasing efficiency for track reconstruction in this region. \par

Several comparisons between data and simulated dijet distributions are presented in Fig.~\ref{fig:Dijet_Mass_DataSimu}. The left panel shows the dijet invariant mass spectrum for all accepted events. The differences in pseudorapidity and azimuthal angle between the two jets (opening angle) only for those events in which one jet is detected in the Barrel and the other in the Endcap are presented on the right.\par

Dijets were also reconstructed in simulation at the particle and parton level, again using the anti-$k_T$ algorithm \cite{Cacciari:2008gp}. As noted previously, particle-level dijets were formed from all stable final-state particles, including those which arise from the underlying event and beam remnants. The parton-level dijets were reconstructed from the hard-scattered partons emitted in the collision, including initial and final-state radiation, but not beam remnants or underlying event effects. Since the detector performance is irrelevant for these jets, the neutral fraction cut and the $p_T$ balance cut were not applied when selecting dijets at the particle or parton levels from the full unbiased \textsc{Pythia} sample.\par

For some systematic uncertainty estimates, it was important to be able to match dijets reconstructed at the particle and parton levels to the `same' simulated dijets reconstructed at the detector level. In practice, we would first find a dijet at the detector level; particle and parton level dijets would then be associated with this dijet if both jets match within $\Delta R = \sqrt{\Delta\eta^2 + \Delta\phi^2} < 0.5$.\par

\section{Experimental Methods}
\subsection{Underlying Event Corrections}
Events with hard jets are often accompanied by a more diffuse background of relatively soft particles. These particles are unrelated to the hard partonic scattering of interest, yet may contribute additional energy and transverse momentum to the reconstructed jets. For the present analysis, primary sources of background are particles generated in the underlying event or from pileup. The former is due to soft processes involving the beam remnants, that is, other partons from the same colliding proton pair, while pileup refers to particles arising from processes that occur at or near the same time as the hard scattering, but that originate from other (usually) $pp$ collisions.\par

For many physics applications, it is useful to estimate the characteristics of these background processes on an event-by-event basis and correct the hard jet kinematics for the effects of the soft contamination. In this analysis, the underlying event observables (energy density and mass density) are constructed for each jet, using the same particle list as that is used as input to the jet finder. This method was developed for the STAR 2012 inclusive jet analysis at $\sqrt{s}$ = 510~GeV~\cite{zchang2016}, and was adapted from the perpendicular cones method used in the ALICE experiment \cite{ALICE:2014dla}.\par

In this method, two cones are defined for the reconstructed jet, each of which is centered at the same $\eta$ as the jet, but rotated $\pm 90^\circ$ away in $\phi$. All particles falling within the two cones are collected. The off-axis cone radius is also chosen to be the same as the jet resolution parameter of the anti-$k_T$ algorithm used in this analysis, $R$ = 0.6. The transverse momentum of each off-axis cone is defined as the scalar sum of the $p_T$ of all the particles inside the cone, and is denoted as $p_{T,ue}$. Similarly, the mass of the off-axis cone is the invariant mass of the vector sum of all the particles inside the cone. The cone transverse momentum density, $\rho_{p_{T},cone}$, is then defined as $p_{T,ue}$ divided by the cone area, $\pi R^2$. The cone mass density, $\rho_{m,cone}$ is the off-axis cone mass divided by the same area. Finally, the underlying event density (transverse momentum or mass) is taken to be the average density of the two cones.\par

The soft background particles of the underlying event are assumed to be evenly distributed over $\eta$-$\phi$ space, so the actual underlying event energy density is expected to be approximately uniform. In practice, though, detector acceptance and efficiency are usually not constant throughout $\eta$-$\phi$ space. The STAR TPC and  electromagnetic calorimetry have very good four-fold symmetry in $\phi$, but not in $\eta$, especially in the forward Endcap region. It is because of these large variations in detector performance with $\eta$ that we chose to evaluate the underlying event densities at the same $\eta$ as that of the jet under consideration, but at values of $\phi$ which should be far from either of the two hard jets in the event.\par

\begin{figure}[!hbt]
    \includegraphics[width=1.0\columnwidth]{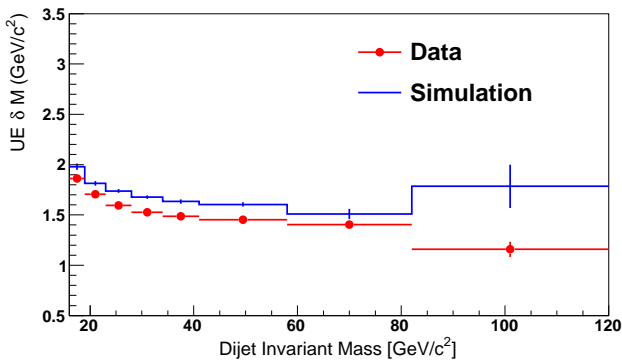}
    \caption{Data/simulation comparisons of the underlying event $\delta M$ (difference before and after the underlying event subtraction) vs. underlying event corrected dijet invariant mass. The points represent the data and the histogram is the simulation.}
    \label{fig:dijet_UE_Mass}
\end{figure}

Dijet measurements are sensitive to both the direction and the mass of each jet, so in general one should always correct the full jet 4-momentum. In this analysis, the underlying event subtraction was performed for each jet using the 4-vector subtraction method from the FastJet group \cite{Cacciari:2011ma}. The equation used is:
\begin{equation}
P^{\mu}_{jet,corr} = P^{\mu}_{jet} - [\rho A_{jet}^{x}, \rho A_{jet}^{y}, (\rho+\rho_{m}) A_{jet}^{z}, (\rho+\rho_{m}) A_{jet}^{E}]
\end{equation}
where $P^{\mu}_{jet}$ is the jet's initial 4-momentum vector, and $P^{\mu}_{jet,corr}$ is the corrected 4-momentum vector after underlying event subtraction; $\rho$ and $\rho_m$ are the average underlying event transverse momentum and mass densities, respectively; and $A_{\mu}$ is the 4-momentum vector area, as calculated by the Fastjet package~\cite{Cacciari:2011ma} using the ghost particle technique~\cite{Cacciari:2007fd}. The underlying event systematic uncertainty was estimated as the difference between data and simulation corrections for the underlying event contribution to the dijet invariant mass as shown in Fig.~\ref{fig:dijet_UE_Mass}.\par
\subsection{Techniques Specific to Endcap Jets}

\subsubsection{Challenges in the Forward (EEMC) Region}
The STAR TPC remains efficient over the nominal range $|\eta| \le 1.3$, but the tracking efficiency decreases rapidly in more forward regions, where much of the Endcap calorimeter is located. Lower tracking efficiency means that jets in the Endcap will be reconstructed at lower $p_{T}$, on average. This effect is seen clearly in simulation, as shown in the upper plot of Fig.~\ref{fig:Jet_ShiftPt}, where the ratio of particle-level jet $p_{T}$ to detector-level $p_{T}$ is plotted as a function of detector $\eta$. This systematic underestimation of the jet $p_{T}$ skews the extraction of the initial state parton momenta. Moreover, jets with a high percentage of neutral energy are preferentially selected over those with most of their energy distributed in charged particles, both in terms of the trigger and jet reconstruction efficiency, leading to a biased sample. The jet mass is also skewed during jet reconstruction. As indicated before, in the jetfinder algorithm tracks are assigned the mass of charged pions, while for the calorimeter towers the particles are assumed to be massless. Both assumptions tend to lower the detector-level jet invariant mass relative to its true value.\par

\subsubsection{Machine Learning Approaches and Corrections}
The Multilayer Perceptron (MLP from TMVA \cite{Hocker:2007ht}), a machine learning regression method, was used to correct the jet $p_{T}$ and mass determined by the jet finding algorithm. Supervised machine-learning regression algorithms make use of training events, for which the desired output is known, to determine an approximation of the underlying functional behavior defining the target value.\par
\begin{figure}[!ht]
  \centering
    \includegraphics[width=1.0\columnwidth]{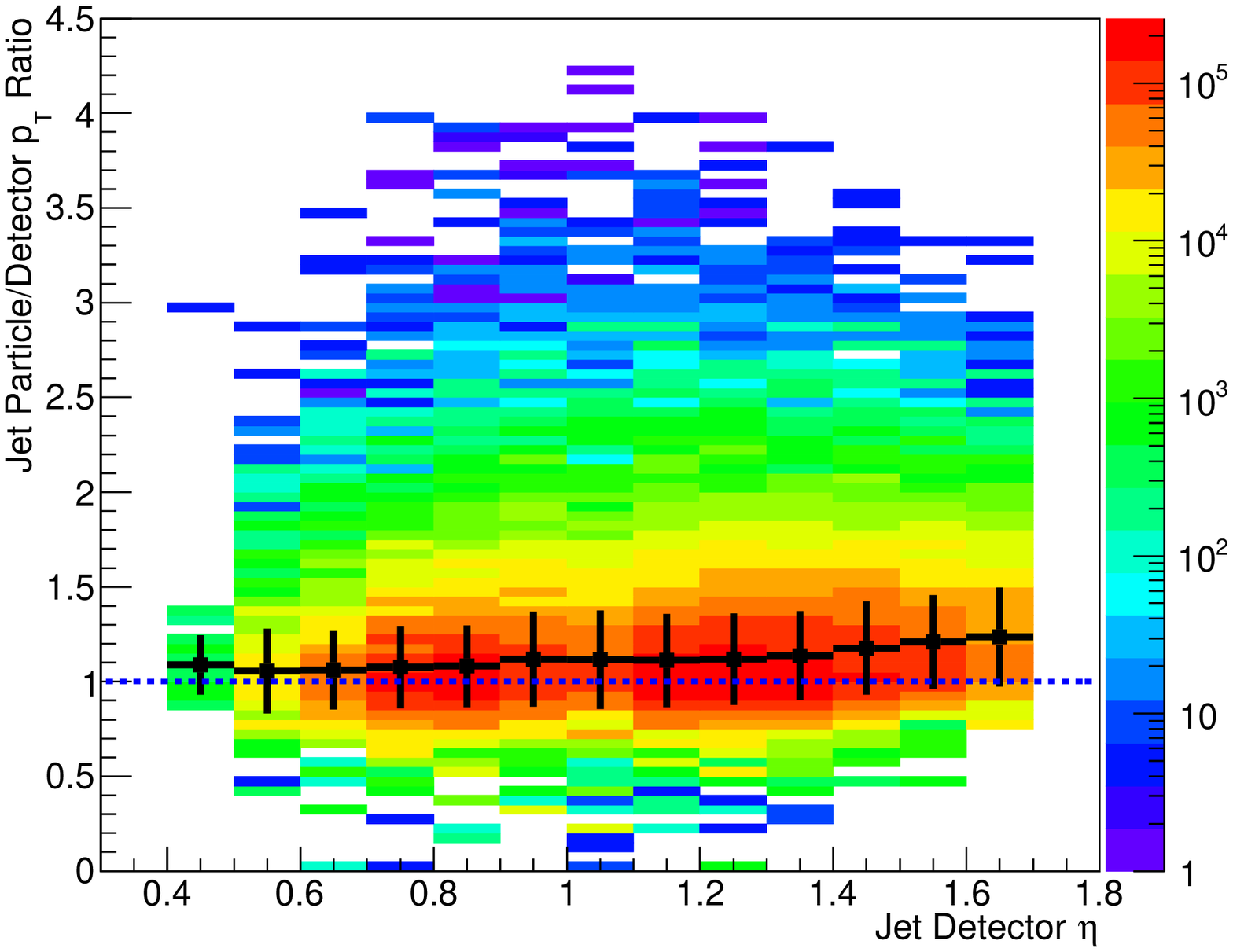}
    \includegraphics[width=1.0\columnwidth]{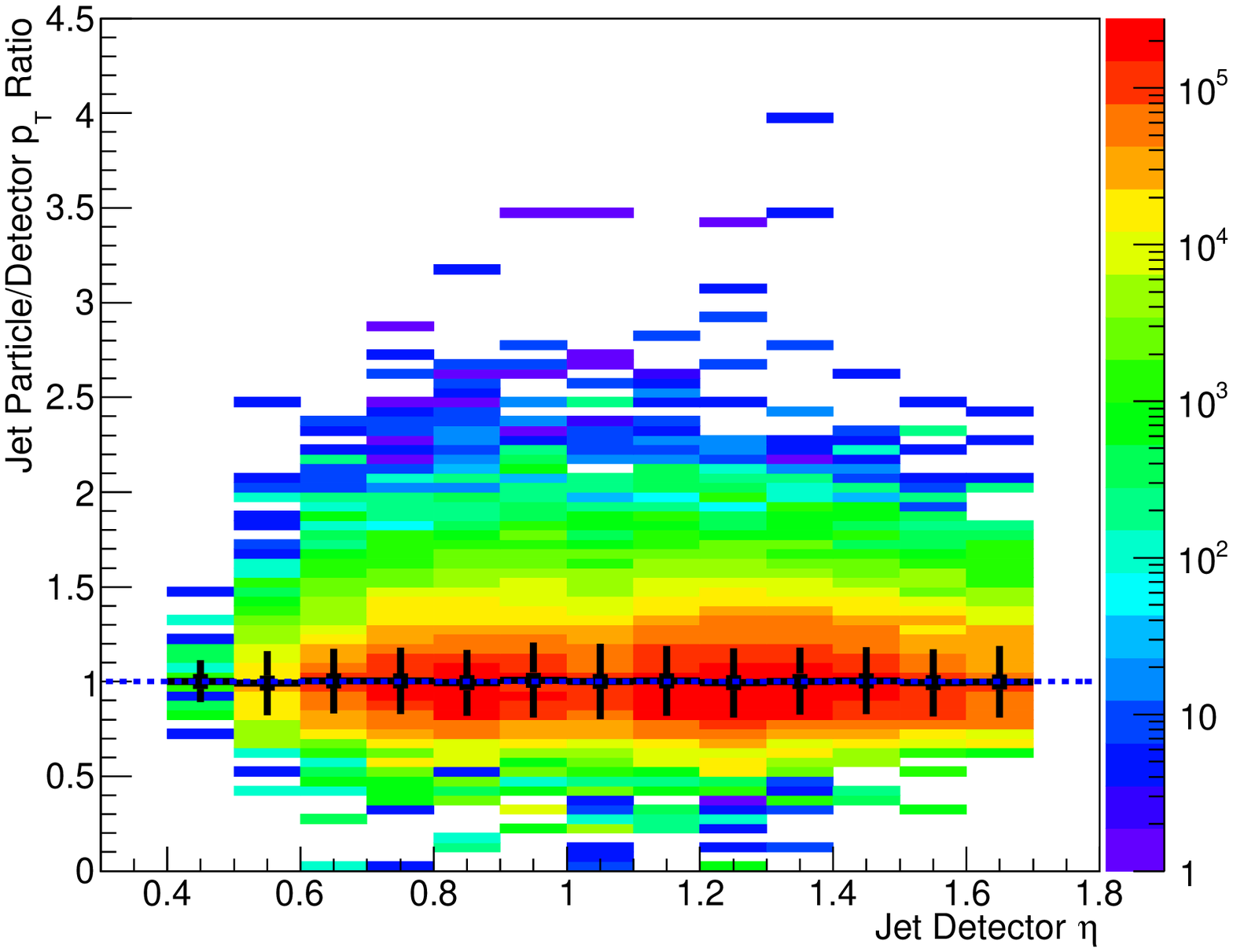}
    \caption{Jet particle-level $p_{T}$ divided by detector-level $p_{T}$ as a function of detector $\eta$ before (upper plot) and after (lower plot) a $p_{T}$ shift correction was made. The correction is determined using machine-learning techniques. The black lines are the average values and the vertical uncertainties are the standard deviations.}
    \label{fig:Jet_ShiftPt}
\end{figure}
\begin{figure}[!ht]
  \centering
    \includegraphics[width=1.00\columnwidth]{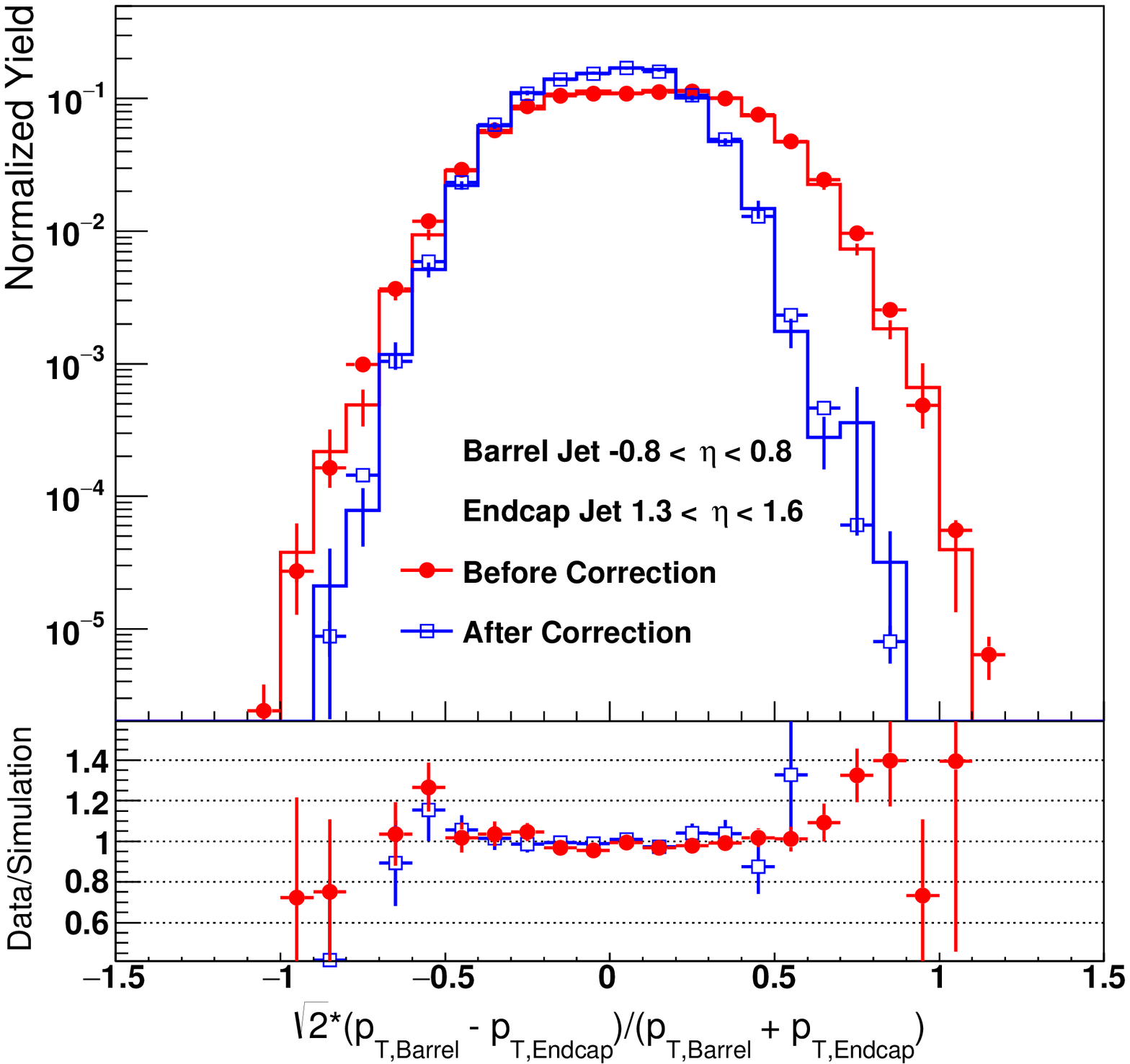}
    \caption{The dijet $p_{T}$ imbalance distribution before (red) and after (blue) $p_T$ corrections were made. The points represent the data (solid circle for before and open square for after the correction), and the histogram is the simulation.}
    \label{fig:dijet_Pt_Imb}
\end{figure}
All of the simulated events that contain Endcap jets were used for the regression study. The key input variables for the jet $p_{T}$ correction are the measured jet $p_{T}$ itself and the detector pseudorapidity. The detector pseudorapidity is used, rather than the jet $\eta$, as it directly corresponds to the detector geometry which affects the tracking efficiency. The jet neutral energy fraction is also used as an input, as it provides information about the bias introduced due to falling tracking efficiency. In addition, the two jets that make up a dijet should have approximately equal transverse momenta, so when correcting the $p_{T}$ of the Endcap jet, the $p_{T}$ of the away-side Barrel jet, which is reconstructed more accurately, is also included as an input to the regression analysis. As noted before, the particle-level to detector-level jet association is performed by looping over all particle-level jets, then selecting the one which is closest in $\eta$-$\phi$ space. The geometric matching condition is that this distance must be less than 0.5. The target value for the jet $p_{T}$ correction is the particle-level jet $p_T$.\par

Using the above method, the network was trained and the associated parameters in the algorithm were optimized. A comparison of the learning output and the target values (particle-level jet $p_T$ over corrected detector-level) is shown in the lower panel of Fig.~\ref{fig:Jet_ShiftPt}. After the machine learning correction is applied, the ratio of particle to detector level jet $p_{T}$ is flat as a function of detector pseudorapidity as seen by the points with uncertainty bars that represent the average of the ratio for each bin. Moreover, the vertical spread in the distribution is also reduced. On average, the resolution of the jet transverse momentum was improved by about $34\%$.\par

Jet $p_{T}$ corrections were also made for the Barrel jets. Though the jet transverse momentum is typically reconstructed more accurately in the Barrel than in the Endcap, the measured $p_T$ is still systematically lower than its true value due the limits on detector performance. For example, the TPC track reconstruction efficiency is estimated to be only $\sim 85\%$ for $|\eta| \le 1.0$. The correction method used for the Barrel jets is identical to that described above, except that the correlated jet $p_{T}$ from the other (Endcap) jet is not used as an input.\par

The net effect of these $p_T$ corrections can be seen in Fig.~\ref{fig:dijet_Pt_Imb}, which shows the dijet $p_T$ imbalance distribution (the difference in magnitude of the two jet $p_T$'s) for events involving Barrel-Endcap dijets. The pseudorapidity of the Endcap jet is required to be between 1.3 and 1.6. Before the correction (red curve), the reconstructed BEMC jet $p_T$ is larger than that of the corresponding EEMC jet on average, and so the distribution is shifted systematically towards positive values. After the correction (blue), the systematic difference is smaller and the spread is also smaller. For the data, the mean value of the distribution changed from 0.086 to -0.009, and the resolution improved by about $40\%$. All of these effects are seen in both the data and in the simulations used to train the method.\par

\begin{figure}[!ht]
  \centering
    \includegraphics[width=1.0\columnwidth]{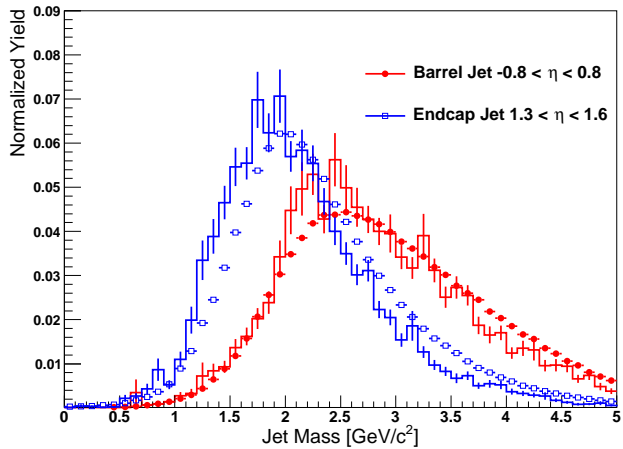}
    \caption{The jet mass distribution after corrections were made. The points represent the data (solid circle for Barrel and open square for Endcap), and the histogram is the simulation.}
    \label{fig:jet_mass_after}
\end{figure}

Even though the jet mass is typically quite small compared to its transverse momentum at RHIC kinematics, it is an important jet property and is needed in calculating the dijet invariant mass. Machine learning techniques were also used to make corrections to the jet mass, following closely the methods described above for jet $p_T$. The input parameters for the artificial neural network were the calculated jet mass, track multiplicity, and tower multiplicity. The falling tracking efficiency also affects the jet mass determination, so the jet transverse momentum, neutral energy fraction and the detector pseudorapidity were also used as input. The target value was the particle-level jet mass from simulation. Figure \ref{fig:jet_mass_after} shows a comparison of the corrected masses for data and simulation. The agreement for Barrel jets is good. The agreement is not as good for Endcap jets. The $\sim$~0.2 GeV/$c^2$ shift between data and simulation in Fig.~\ref{fig:jet_mass_after}  results in a negligible error ($\ll 0.1$ GeV/$c^2$) on the correction to the dijet invariant mass scale.\par

In this analysis, both the $p_{T}$ and mass for the Barrel and Endcap jets were corrected separately, and a dijet invariant mass was calculated using the corrected jet transverse momentum and mass from machine learning. The dijet invariant mass was found by taking the square of the sum of the 4-momenta of the two jets which make up the dijet:
\begin{equation}
  M^{2}_{3,4} = (P_{3} + P_{4})^{2}
\end{equation}
\begin{widetext}
\begin{equation}
  M_{3,4} = \sqrt{m^{2}_{3} + m^{2}_{4} + 2\sqrt{m^{2}_{3} + p^{2}_{T3}}\sqrt{m^{2}_{4} + p^{2}_{T4}}\cosh(\Delta y) - 2p_{T3}p_{T4}\cos(\Delta \phi)}.
\label{dijet_Mass_Equa}
\end{equation}
\end{widetext}
where $m$ and $p_{T}$ are the mass and transverse momentum of the jet, $\Delta y$ is the rapidity difference and $\Delta \phi$ is the $\phi$ difference of the two jets.\par
\section{The Spin Asymmetry $A_{LL}$}
The spin observable measurable at RHIC that is most directly sensitive to the helicities of gluons within the proton, $\Delta g(x)$, is the longitudinal double-spin asymmetry $A_{LL}$. STAR has published $A_{LL}$ measurements for inclusive jet \cite{Abelev:2006uq, Abelev:2007vt, Adamczyk:2012qj, Adamczyk:2015}, mid-rapidity dijet \cite{Adamczyk:2016okk}, mid-rapidity $\pi^{0}$ \cite{Abelev:2009pb}, intermediate rapidity $\pi^{0}$ \cite{Adamczyk:2013yvv} and forward rapidity $\pi^{0}$ final states \cite{FMS:2018}. Taken together, these results have placed strong constraints on our current understanding of the gluon helicity distribution.\par

The longitudinal double-spin asymmetry $A_{LL}$ is defined in terms of helicity-dependent cross sections:
\begin{equation}
A_{LL} \equiv \frac{\sigma_{++} - \sigma_{+-}}{\sigma_{++} + \sigma_{+-}}
\label{equ:ALL_cross}
\end{equation}
where $\sigma_{++}$ and $\sigma_{+-}$ are the differential production cross sections when the beam protons have equal and opposite helicities, respectively. Experimentally, sorting the measured yields by beam spin state, and combining many independent measurements, enables a precise determination of $A_{LL}$. In practice, the asymmetry is evaluated as:
\begin{equation}
A_{LL} = \frac{\sum(P_{Y}P_{B})(N^{++}-rN^{+-})}{\sum(P_{Y}P_{B})^{2}(N^{++}+rN^{+-})},
\label{equ:ALL}
\end{equation}
where $P_{Y,B}$ are the measured polarizations of the Yellow and Blue beams, $N^{++}$ and $N^{+-}$ are the dijet yields from proton beam bunches with equal and opposite helicity configurations. The relative luminosity, $r$, was calculated from the observed bunch-by-bunch BBC coincidence rates after corrections for accidental and multiple hits. The sum in Eq.~\ref{equ:ALL} is over individual data runs, which in 2009 ranged from 10 to 60 minutes in length. It is important to note that these run lengths are quite short compared to the time scales over which the beam polarizations and relative luminosities were observed to vary.\par

\subsection{Beam Polarizations}
The beams are not $100\%$ polarized, so the measured asymmetries need to be scaled by the beam polarizations, as indicated in Eq.~\ref{equ:ALL}. The general scheme used for polarization measurements was discussed in Sec.\ III.A; here we focus on the individual run information~\cite{RHICPolG}. For each fill, the RHIC polarimetry group provided a luminosity-weighted polarization for each beam, as well as an initial polarization and a value for the change in polarization over time. In order to account for polarization loss over time, the value of the polarization was determined from the Unix timestamp $t$ of each run using the equation:
\begin{equation}
P(t) = P_{0} + \frac{dP}{dt}(t-t_{0})
\end{equation}
where $P_{0}$ is the initial polarization, slope $\frac{dP}{dt}$ is the polarization change with time, and $t_{0}$ is the Unix start time of the fill.\par

The reason for adopting the event-time-dependent polarizations described above is due to the STAR trigger optimization algorithm. The average polarization value reported by the RHIC polarimetry group for each fill was weighted by the luminosity over the course of that fill. Thus, if the rate at which events are recorded scales proportionally with the instantaneous luminosity, the average polarization would be the correct value to use. This proportionality roughly holds for the JP2 events, as that trigger was not prescaled throughout the run. The JP1 trigger, however, was prescaled, and the prescale value was chosen to match the available trigger bandwidth at the beginning of each run during a fill. Since the luminosity drops significantly over the course of a fill, along with the rates of non-prescaled triggers, the JP1 events are always acquired at a higher rate near the end of a fill. Using the fill-averaged polarization value for the JP1 sample would thus tend to overestimate the beam polarizations appropriate for this sample; calculating $A_{LL}$ using the beam polarizations found as a function of event time alleviates this problem.\par

\subsection{Relative luminosity}
As shown in Eq. \ref{equ:ALL}, extraction of $A_{LL}$ also requires precise knowledge of the ratio of integrated luminosities between the two beam spin states, but absolute luminosities are not needed. However, there are only a limited number of bunch crossings available in the collider, and not all bunches have the same intensity, so some spin state combinations may sample more luminosity than others. Therefore, each yield must be normalized by the associated luminosity. The bunch-by-bunch spin patterns used when filling the RHIC rings, and details of calculating the relative luminosity ratios, are constructed in such a way to cancel out many sources of false asymmetries \cite{Adamczyk:2012qj} which would distort the value of $r$.\par

\section{Biases and Corrections}
\subsection{Dijet Invariant Mass Correction}
In order to compare our experimental results with theoretical predictions, which are calculated at the parton level, a determination of the parton-level dijet invariant mass of each data point was made by applying a simple mass shift to each point. This mass correction accounts for the difference in parton and particle-level dijet invariant mass scales. The machine learning procedure described in the previous section corrects jets back to the particle level, so this additional mass shift is found by comparing the particle-level masses to the matched parton-level dijet masses. For a given particle-level mass bin, the difference between the parton and particle-level dijet masses was calculated event-by-event. The correction was then taken as the mean value of these differences, averaged over the entire event sample. The final data points are plotted at this average particle-level mass, plus the particle-to-parton estimated mass shift as shown in Table~\ref{table:Dijet_Correction_BE}.\par

\subsection{Trigger and Reconstruction bias}
The values of $A_{LL}$ extracted from the data represent an admixture of the asymmetries produced from the three dominant partonic scattering sub-processes: quark-quark ($qq$), quark-gluon ($qg$), and gluon-gluon ($gg$). The STAR jet-patch trigger may be more efficient for certain sub-processes, which will alter the sub-process fractions in the data sample compared to the physically correct fractions, thereby shifting the measured $A_{LL}$. Further distortions can arise due to systematic shifts caused by the finite resolution of the detector, coupled with a rapidly falling invariant mass distribution, and thus change the sub-process fraction associated with a given mass. A trigger and reconstruction bias correction was applied to the raw $A_{LL}$ values to compensate for these effects.\par

In order to determine the bias introduced by the trigger and jet reconstruction methods, polarized PDF's, which are not well known, are needed, in addition to the more tightly constrained unpolarized PDF's. The NNPDFPol1.1 PDF set \cite{Nocera:2014gqa} was used as input, as the best-fit values agree well with STAR results, and the publicly available replica sets provide a robust way to determine the uncertainty on the correction. Parameterizations of the polarized parton distribution functions are combined with \textsc{Pythia} parton kinematic variables to generate predictions of $A_{LL}$ vs. dijet mass for a particular model at both the parton and detector levels.\par

The trigger and reconstruction bias correction for each mass bin was calculated by evaluating the quantity
\begin{equation}
\Delta A_{LL} = A^{Det}_{LL} - A^{parton}_{LL}
\end{equation}
for each of the 100 replica NNPDF sets, where $A^{Det}_{LL}$ is the $A_{LL}$ value found for detector-level dijets in the simulation and $A^{parton}_{LL}$ is the $A_{LL}$ value found for parton-level dijets, calculated at the average parton-level dijet mass that is sampled by the detector dijet bin. The correction was taken to be the average of the 100 values for $\Delta A_{LL}$ calculated; the final result is then $A_{LL}^{final} = A_{LL}^{raw} - \Delta A_{LL}^{average}$. The statistical uncertainties of the detector-level NNPDF $A_{LL}$ and the square root of the variance of the 100 $\Delta A_{LL}$ were added in quadrature, and were assigned as the systematic uncertainty on dijet $\Delta A_{LL}$. Final values of these quantities for events with different dijet topologies are shown in Table~\ref{table:Dijet_Correction_BE}.\par

\begin{table*}[!htb]
  \setlength{\tabcolsep}{6pt}
  \centering
  \begin{tabular}{c c c c c}
\hline \hline
  \multicolumn{5}{c}{East Barrel-Endcap}\\
    & Detector Level & Particle Level & Particle to Parton &  \\
Bin & Mass Range (GeV/$c^{2}$) & Ave Mass (GeV/$c^{2}$) & Mass Shift (GeV/$c^{2}$) & Trigger and Reco Shift\\
\hline
1 & 16 - 19 & 18.07 & 0.37 $\pm$ 0.40 & 0.0005 $\pm$ 0.0006 \\
2 & 19 - 23 & 21.22 & 0.90 $\pm$ 0.14 & 0.0006 $\pm$ 0.0004 \\
3 & 23 - 28 & 25.41 & 1.17 $\pm$ 0.17 & 0.0012 $\pm$ 0.0004 \\
4 & 28 - 34 & 30.68 & 1.54 $\pm$ 0.11 & 0.0010 $\pm$ 0.0008 \\
5 & 34 - 41 & 36.95 & 1.40 $\pm$ 0.14 & 0.0016 $\pm$ 0.0010 \\
6 & 41 - 58 & 46.24 & 1.77 $\pm$ 0.14 & 0.0019 $\pm$ 0.0010 \\
7 & 58 - 82 & 63.84 & 1.89 $\pm$ 0.34 & 0.0069 $\pm$ 0.0060 \\
\hline \hline
  \multicolumn{5}{c}{West Barrel-Endcap}\\
    & Detector Level & Particle Level & Particle to Parton &  \\
Bin & Mass Range (GeV/$c^{2}$) & Ave Mass (GeV/$c^{2}$) & Mass Shift (GeV/$c^{2}$) & Trigger and Reco Shift\\
\hline
1 & 16 - 19 & 17.68 & 0.83 $\pm$ 0.11 & 0.0005 $\pm$ 0.0006 \\
2 & 19 - 23 & 20.93 & 0.85 $\pm$ 0.09 & 0.0006 $\pm$ 0.0005 \\
3 & 23 - 28 & 25.22 & 0.80 $\pm$ 0.14 & -0.0001 $\pm$ 0.0004 \\
4 & 28 - 34 & 30.47 & 0.32 $\pm$ 0.72 & 0.0001 $\pm$ 0.0009 \\
5 & 34 - 41 & 36.75 & 1.20 $\pm$ 0.12 & -0.0003 $\pm$ 0.0015 \\
6 & 41 - 58 & 45.51 & 0.91 $\pm$ 0.16 & 0.0023 $\pm$ 0.0026 \\
7 & 58 - 82 & 62.57 & 0.26 $\pm$ 0.66 & -0.0078 $\pm$ 0.0056 \\
\hline \hline
  \multicolumn{5}{c}{Endcap-Endcap}\\
    & Detector Level & Particle Level &  Particle to Parton &  \\
Bin & Mass Range (GeV/$c^{2}$) & Ave Mass (GeV/$c^{2}$) & Mass Shift (GeV/$c^{2}$) & Trigger and Reco Shift\\
\hline
1 & 16 - 19 & 17.54 & 0.96 $\pm$ 0.14 & -0.0002 $\pm$ 0.0008 \\
2 & 19 - 23 & 20.79 & 0.92 $\pm$ 0.15 & -0.0008 $\pm$ 0.0009 \\
3 & 23 - 28 & 24.98 & 1.33 $\pm$ 0.15 & 0.0007 $\pm$ 0.0014 \\
4 & 28 - 34 & 30.17 & 1.57 $\pm$ 0.20 & 0.0006 $\pm$ 0.0031 \\
5 & 34 - 41 & 36.13 & 2.75 $\pm$ 0.39 & 0.0091 $\pm$ 0.0052 \\
\hline \hline
  \end{tabular}
  \caption{Dijet parton-level corrections for different event topologies}
  \label{table:Dijet_Correction_BE}
\end{table*}

\subsection{Systematic Uncertainty Estimates}
The systematic uncertainties were divided into two categories: systematic uncertainty on the calculated dijet invariant mass (``x-axis uncertainties") and those on the actual $A_{LL}$ asymmetries (``y-axis uncertainties"). The systematic uncertainty on $A_{LL}$ includes the beam polarization uncertainty, the relative luminosity uncertainty, the underlying event systematic uncertainty, the trigger and reconstruction bias uncertainty, and the residual transverse polarization uncertainty. Systematic uncertainties on the dijet invariant mass include the jet energy scale uncertainty, tracking efficiency uncertainty, jet $p_T$ and mass correction uncertainties, the dijet invariant mass shift uncertainties, and uncertainties associated with the choice of \textsc{Pythia} tune. Some of these have been described in previous sections, while others merit more discussion below.\par

The uncertainty in the product of the average beam polarizations (the relevant quantity for double-spin asymmetries) was determined by the RHIC polarimetry group, and was estimated to be $6.5\%$~\cite{RHICPolG}. The relative luminosity systematic is the same as that determined for the inclusive jet and mid-rapidity dijet analyses ($\pm 0.0005$), which applies to all the mass bins. This was determined by examining BBC/ZDC differences \cite{Kiryluk:2005gg, Adler:2000bd} and evaluating a number of ``false" single and double-spin asymmetries which are expected to yield null results.\par

A complete list of the final results on dijet invariant mass systematic uncertainties for the different dijet topologies is shown in Table \ref{table:Dijet_MassSystematic_FullTopo}. Table \ref{table:Dijet_ALLSystematic_FullTopo} is the equivalent table for systematic uncertainties on $A_{LL}$.\par

\subsubsection{Dijet Energy Scale Systematic Uncertainties}
A significant source of systematic uncertainty on the reconstructed dijet mass comes from the jet energy scale uncertainty. The jet energy scale uncertainties consist of two parts: one from the scale and status uncertainties of the EMC towers, and the other from the TPC track transverse momentum uncertainty and the uncertainty in the tower response to charged hadrons. Contributions from the $\eta$-$\phi$ position uncertainties for individual jets are negligible and are not considered in this analysis.\par

The BEMC scale uncertainty was estimated to be $4.6\%$ while the EEMC scale uncertainty is $4.5\%$ \cite{TingLin2017}. The BEMC and EEMC status uncertainties, {\it i.e.}, how well the monitoring software kept up with failed channels, were estimated at $1\%$. EMC tower-track response uncertainty was taken as $2.5\%$ for jets measured in Barrel and $2.3\%$ for jets measured in the Endcap \cite{Adams:2004cb} \cite{HuoLiaoyuan2012}. The final dijet energy scale uncertainties are shown in the third column of Table~\ref{table:Dijet_MassSystematic_FullTopo}.\par

Effects due to uncertainties in the tracking efficiency were calculated by comparing the average dijet invariant mass difference between detector and parton level using the full set of reconstructed tracks of the TPC, against the same quantity when using only a partial set of reconstructed tracks. The partial set of reconstructed tracks from the TPC was chosen by randomly rejecting a certain percent of tracks from the full set before performing jet reconstruction. In this analysis, the rejection fraction was chosen to be $7\%$. This is larger than the typical STAR tracking efficiency uncertainty because the short tracks at $\eta > 1$ provide much less determination. The values determined are shown in the fourth column of Table~\ref{table:Dijet_MassSystematic_FullTopo}.\par

Systematic uncertainties on the dijet invariant mass shift also include the uncertainties which arise due to the limited statistics of the simulation sample. The statistical uncertainty was determined by adding in quadrature the uncertainties from the various trigger samples, weighted by the trigger fractions. The final values are shown in the fifth column of Table~\ref{table:Dijet_MassSystematic_FullTopo}.\par

Finally, the dijet invariant mass systematic uncertainties due to the underlying event processes were calculated by taking the difference of the underlying event contributions to the dijet mass found between estimates derived from data vs.\ those determined using the embedding sample. These uncertainties are shown in the seventh column of Table~\ref{table:Dijet_MassSystematic_FullTopo}.\par

\subsubsection{\textsc{Pythia} Tune Systematic Uncertainties}
\textsc{Pythia} parameters can be varied independently to fit various data sets. There are also several `standard' tune sets available. The dijet invariant mass correction uncertainties due to the choice of \textsc{Pythia} tune were estimated in this analysis by utilizing the possible variants provided for Perugia0 in the \textsc{Pythia} version of 6.4.26 (tune 320 to 328) and Perugia2012 in \textsc{Pythia} 6.4.28 \cite{Skands:2010ak}. The invariant mass shifts between the particle-level dijet and parton-level dijet were calculated, and the differences between those shifts were used as the \textsc{Pythia} tune systematic uncertainties. We note that tune 328 would include an alternate dependence on underlying event contributions. It might result in double counting the underlying event uncertainties that have already been estimated from the data vs. simulation difference, so tune 328 was not used here. In addition, tunes 321 and 322 vary the same parameters in opposite directions, so half of the absolute difference between the two results was used. The quadrature sum of the differences among the shifts resulting from using different tune sets was taken as the final uncertainty estimate, and is shown in the eighth column of Table~\ref{table:Dijet_MassSystematic_FullTopo}.\par

\setlength{\tabcolsep}{6pt}
\begin{table*}[!htb]
  \centering
  \begin{tabular}{ccccccccc}
\hline \hline
  \multicolumn{9}{c}{East Barrel-Endcap}\\
Bin & Ave Mass & Energy Scale & Tracking Eff. & Mass Shift & Machine Learning & UE Syst. & Tune Syst. & Total \\
\hline
1 & 18.44 & 0.53 & 0.28 & 0.40 & 0.16 & 0.22 & 1.15 & 1.38 \\
2 & 22.11 & 0.64 & 0.26 & 0.14 & 0.14 & 0.04 & 0.68 & 0.99 \\
3 & 26.58 & 0.77 & 0.32 & 0.17 & 0.10 & 0.05 & 0.79 & 1.16 \\
4 & 32.21 & 0.92 & 0.28 & 0.11 & 0.08 & 0.03 & 1.20 & 1.55 \\
5 & 38.35 & 1.09 & 0.43 & 0.14 & 0.14 & 0.09 & 0.72 & 1.40 \\
6 & 48.01 & 1.36 & 0.47 & 0.14 & 0.26 & 0.11 & 0.72 & 1.65 \\
7 & 65.73 & 1.86 & 0.54 & 0.34 & 0.51 & 0.13 & 0.69 & 2.15 \\
\hline \hline
  \multicolumn{9}{c}{West Barrel-Endcap}\\
Bin & Ave Mass & Energy Scale & Tracking Eff. & Mass Shift & Machine Learning & UE Syst. & Tune Syst. & Total \\
\hline
1 & 18.51 & 0.53 & 0.23 & 0.11 & 0.12 & 0.15 & 1.02 & 1.20 \\
2 & 21.78 & 0.63 & 0.33 & 0.09 & 0.08 & 0.08 & 0.91 & 1.16 \\
3 & 26.02 & 0.75 & 0.26 & 0.14 & 0.08 & 0.02 & 0.87 & 1.19 \\
4 & 30.79 & 0.88 & 0.30 & 0.72 & 0.10 & 0.08 & 0.68 & 1.37 \\
5 & 37.96 & 1.08 & 0.35 & 0.12 & 0.21 & 0.08 & 0.64 & 1.33 \\
6 & 46.43 & 1.32 & 0.38 & 0.16 & 0.55 & 0.12 & 0.41 & 1.55 \\
7 & 62.82 & 1.79 & 0.21 & 0.67 & 2.52 & 0.01 & 0.56 & 3.22 \\
\hline \hline
  \multicolumn{9}{c}{Endcap-Endcap}\\
Bin & Ave Mass & Energy Scale & Tracking Eff. & Mass Shift & Machine Learning & UE Syst. & Tune Syst. & Total \\
\hline
1 & 18.50 & 0.64 & 0.18 & 0.14 & 0.16 & 0.17 & 0.50 & 0.88 \\
2 & 21.70 & 0.76 & 0.12 & 0.15 & 0.13 & 0.05 & 0.90 & 1.20 \\
3 & 26.31 & 0.91 & 0.21 & 0.15 & 0.11 & 0.06 & 0.86 & 1.28 \\
4 & 31.74 & 1.10 & 0.25 & 0.20 & 0.08 & 0.08 & 1.31 & 1.74 \\
5 & 38.88 & 1.31 & 0.36 & 0.39 & 0.09 & 0.09 & 0.52 & 1.51 \\
\hline \hline
  \end{tabular}
  \caption{Systematics Uncertainties on dijet invariant mass for (GeV/$c^{2}$) the different event topologies}
  \label{table:Dijet_MassSystematic_FullTopo}
\end{table*}

\subsubsection{Systematic Uncertainties on Machine Learning Correction}
Some machine learning techniques adapted in this analysis, such as Multilayer Perceptron method, may be sensitive to the network parameters’ change. In the Multilayer Perceptron method, for example, small changes to the network parameters, such as the number of layers or nodes, may impact the learning process. Alternate machine-learning algorithms will also determine corrections slightly differently. To account for these effects, systematic uncertainties for the jet $p_{T}$ and mass corrections were evaluated by comparing the output from slightly modified input and network parameter sets, or by using alternate methods, with the differences added in quadrature. For the Multilayer Perceptron, the training sample size, number of layers, and number of nodes were systematically varied. To test sensitivity to the choice of algorithm, the Linear Discriminant (LD from TMVA) and K-Nearest Neighbors (KNN from TMVA) packages were used as alternate methods. The final uncertainty is shown in the sixth column of Table~\ref{table:Dijet_MassSystematic_FullTopo}.\par

\subsubsection{Residual Transverse Beam Polarization}
Due to imperfect tuning of the spin rotators in the collider, each beam polarization direction may be left with a residual transverse component. The resulting contribution to $A_{LL}$ can be evaluated as
\begin{equation}
  \delta A_{LL} = |\tan \theta_{Y}\tan \theta_{B}\cos(\phi_{Y} - \phi_{B})A_{\Sigma}| 
  \label{equ:residualALL}
\end{equation}
where $\theta$ and $\phi$ are the polar and azimuthal angles of the polarization  directions for the Yellow and Blue beams, and $A_{\Sigma}$ is the relevant transverse spin asymmetry. The correction method employed here is similar to what has been done in previous inclusive jet analyses at STAR \cite{Adamczyk:2015}. Since there was no dedicated transverse running during 2009, the $A_{\Sigma}$ values used were those measured in 2006 \cite{Adamczyk:2012qj}. These values were all consistent with zero, so the statistical uncertainty on the $A_{\Sigma}$ measurement was taken and used in the calculation of the systematic uncertainty. To simplify the calculation and set an upper limit on the systematic, the $\cos(\phi_{Y} - \phi_{B})$ term was set to 1. The transverse residual double-spin asymmetry uncertainty was found to be of the same order of magnitude as the relative luminosity uncertainty, and the values are shown in the third column of Table~\ref{table:Dijet_ALLSystematic_FullTopo}.\par

\subsubsection{Underlying Event Systematic Uncertainties on $A_{LL}$}
The contributions of the underlying event to the dijet invariant mass were discussed in Sec.\ V.A. In addition, if $\delta M$ has a longitudinal double-spin dependence, it can introduce an apparent mass shift between dijets in like and unlike helicity collisions, thereby producing a systematic error in the dijet $A_{LL}$. The measured $\delta M$ values were examined for spin dependence.  No effect was found; upper limits of $<0.2$\% for Barrel-Endcap dijets and $<0.4$\% for Endcap-Endcap dijets were established.  The limits were then used to estimate changes of the dijet cross section due to the underlying events, which were assigned as the corresponding systematic uncertainties. The final results are shown in the fourth column of Table~\ref{table:Dijet_ALLSystematic_FullTopo}.\par

\begin{table*}[!htb]
  \setlength{\tabcolsep}{6pt}
  \centering
  \begin{tabular}{cccccc}
\hline \hline
  \multicolumn{5}{c}{East Barrel-Endcap}\\
Bin & Ave Mass (GeV/$c^{2}$) & Trans Residual & UE & Trigger and Reco. & Total \\
\hline
1 & 18.44 & 0.0003 & 0.0007 & 0.0006 & 0.0010 \\
2 & 22.11 & 0.0003 & 0.0014 & 0.0004 & 0.0015 \\
3 & 26.58 & 0.0003 & 0.0016 & 0.0004 & 0.0016 \\
4 & 32.21 & 0.0003 & 0.0010 & 0.0008 & 0.0013 \\
5 & 38.35 & 0.0006 & 0.0013 & 0.0010 & 0.0017 \\
6 & 48.01 & 0.0013 & 0.0022 & 0.0010 & 0.0027 \\
7 & 65.73 & 0.0024 & 0.0022 & 0.0060 & 0.0068 \\
\hline \hline
  \multicolumn{5}{c}{West Barrel-Endcap}\\
Bin & Ave Mass(GeV/$c^{2}$) & Trans Residual & UE & Trigger and Reco. & Total \\
\hline
1 & 18.51 & 0.0003 & 0.0010 & 0.0006 & 0.0012 \\
2 & 21.78 & 0.0003 & 0.0014 & 0.0005 & 0.0015 \\
3 & 26.02 & 0.0003 & 0.0005 & 0.0004 & 0.0007 \\
4 & 30.79 & 0.0003 & 0.0011 & 0.0009 & 0.0015 \\
5 & 37.96 & 0.0006 & 0.0009 & 0.0015 & 0.0019 \\
6 & 46.43 & 0.0012 & 0.0021 & 0.0026 & 0.0036 \\
7 & 62.82 & 0.0022 & 0.0021 & 0.0056 & 0.0064 \\
\hline \hline
  \multicolumn{5}{c}{Endcap-Endcap}\\
Bin & Ave Mass(GeV/$c^{2}$) & Trans Residual & UE & Trigger and Reco. & Total \\
\hline
1 & 18.50 & 0.0003 & 0.0019 & 0.0008 & 0.0020 \\
2 & 21.70 & 0.0003 & 0.0022 & 0.0009 & 0.0024 \\
3 & 26.31 & 0.0003 & 0.0006 & 0.0014 & 0.0016 \\
4 & 31.74 & 0.0003 & 0.0044 & 0.0031 & 0.0054 \\
5 & 38.88 & 0.0004 & 0.0044 & 0.0052 & 0.0068 \\
\hline \hline
  \end{tabular}
  \caption{Systematic uncertainties on $A_{LL}$ for the different dijet topologies}
  \label{table:Dijet_ALLSystematic_FullTopo}
\end{table*}

\section{Spin Asymmetry Results}
\subsection{Experimental results}
Table \ref{table:Dijet_Systematic_Diff} lists our final results for the spin asymmetry $A_{LL}$ at different dijet invariant mass values. The results are separated into three dijet event topologies: dijets in which one jet is detected in the east half of the Barrel EMC ($-0.8 < \eta_{\rm jet} < 0.0$) or in the west half of the Barrel EMC ($0.0 < \eta_{\rm jet} < 0.8$), while the other is in the Endcap ($0.8 < \eta_{\rm jet} < 1.8$); and events in which both jets fall in the Endcap. The correlation matrix between the 2009 inclusive jet $A_{LL}$ measurement \cite{Adamczyk:2015} and these dijet results can be found in the supplemental materials \cite{Supplement}.\par

\begin{table}[!htb]
  \centering
  \begin{tabular}{ccc}
\hline \hline
  \multicolumn{3}{c}{East Barrel-Endcap}\\
Bin   & Mass $\pm$ (Sys) [GeV/$c^2$] & $A_{LL} \pm$ (Stat) $\pm$ (Sys)  \\
\hline
1 & 18.44 $\pm$ 1.38 & -0.0178 $\pm$ 0.0106 $\pm$ 0.0010 \\
2 & 22.11 $\pm$ 0.99 & 0.0058 $\pm$ 0.0047 $\pm$ 0.0015 \\
3 & 26.58 $\pm$ 1.16 & 0.0048 $\pm$ 0.0039 $\pm$ 0.0016 \\
4 & 32.21 $\pm$ 1.55 & 0.0017 $\pm$ 0.0044 $\pm$ 0.0013 \\
5 & 38.35 $\pm$ 1.40 & -0.0078 $\pm$ 0.0061 $\pm$ 0.0017 \\
6 & 48.01 $\pm$ 1.65 & 0.0099 $\pm$ 0.0084 $\pm$ 0.0027 \\
7 & 65.73 $\pm$ 2.15 & 0.0120 $\pm$ 0.0296 $\pm$ 0.0068 \\
\hline
\hline
  \multicolumn{3}{c}{West Barrel-Endcap}\\
Bin    & Mass $\pm$ (Sys) [GeV/$c^2$] & $A_{LL} \pm$ (Stat) $\pm$ (Sys)  \\
\hline
1 & 18.51 $\pm$ 1.20 & -0.0034 $\pm$ 0.0046 $\pm$ 0.0012 \\
2 & 21.78 $\pm$ 1.16 & 0.0131 $\pm$ 0.0036 $\pm$ 0.0015 \\
3 & 26.02 $\pm$ 1.19 & 0.0027 $\pm$ 0.0040 $\pm$ 0.0007 \\
4 & 30.79 $\pm$ 1.37 & 0.0066 $\pm$ 0.0057 $\pm$ 0.0015 \\
5 & 37.96 $\pm$ 1.33 & 0.0209 $\pm$ 0.0095 $\pm$ 0.0019 \\
6 & 46.43 $\pm$ 1.55 & 0.0113 $\pm$ 0.0163 $\pm$ 0.0036 \\
7 & 62.82 $\pm$ 3.22 & 0.0314 $\pm$ 0.0871 $\pm$ 0.0064 \\
\hline
\hline
  \multicolumn{3}{c}{Endcap-Endcap}\\
Bin    & Mass $\pm$ (Sys) [GeV/$c^2$] & $A_{LL} \pm$ (Stat) $\pm$ (Sys)  \\
\hline
1 & 18.50 $\pm$ 0.88 & 0.0019 $\pm$ 0.0069 $\pm$ 0.0020 \\
2 & 21.70 $\pm$ 1.20 & -0.0069 $\pm$ 0.0069 $\pm$ 0.0024 \\
3 & 26.31 $\pm$ 1.28 & 0.0212 $\pm$ 0.0099 $\pm$ 0.0016 \\
4 & 31.74 $\pm$ 1.74 & 0.0425 $\pm$ 0.0190 $\pm$ 0.0054 \\
5 & 38.88 $\pm$ 1.51 & 0.0779 $\pm$ 0.0458 $\pm$ 0.0068 \\ 
\hline \hline
  \end{tabular}
  \caption{Final values and uncertainties for dijet $A_{LL}$ at parton level dijet invariant mass}
  \label{table:Dijet_Systematic_Diff}
\end{table}

\begin{figure*}
\centering
\begin{minipage}{.49\textwidth}
  \includegraphics[width=1\columnwidth]{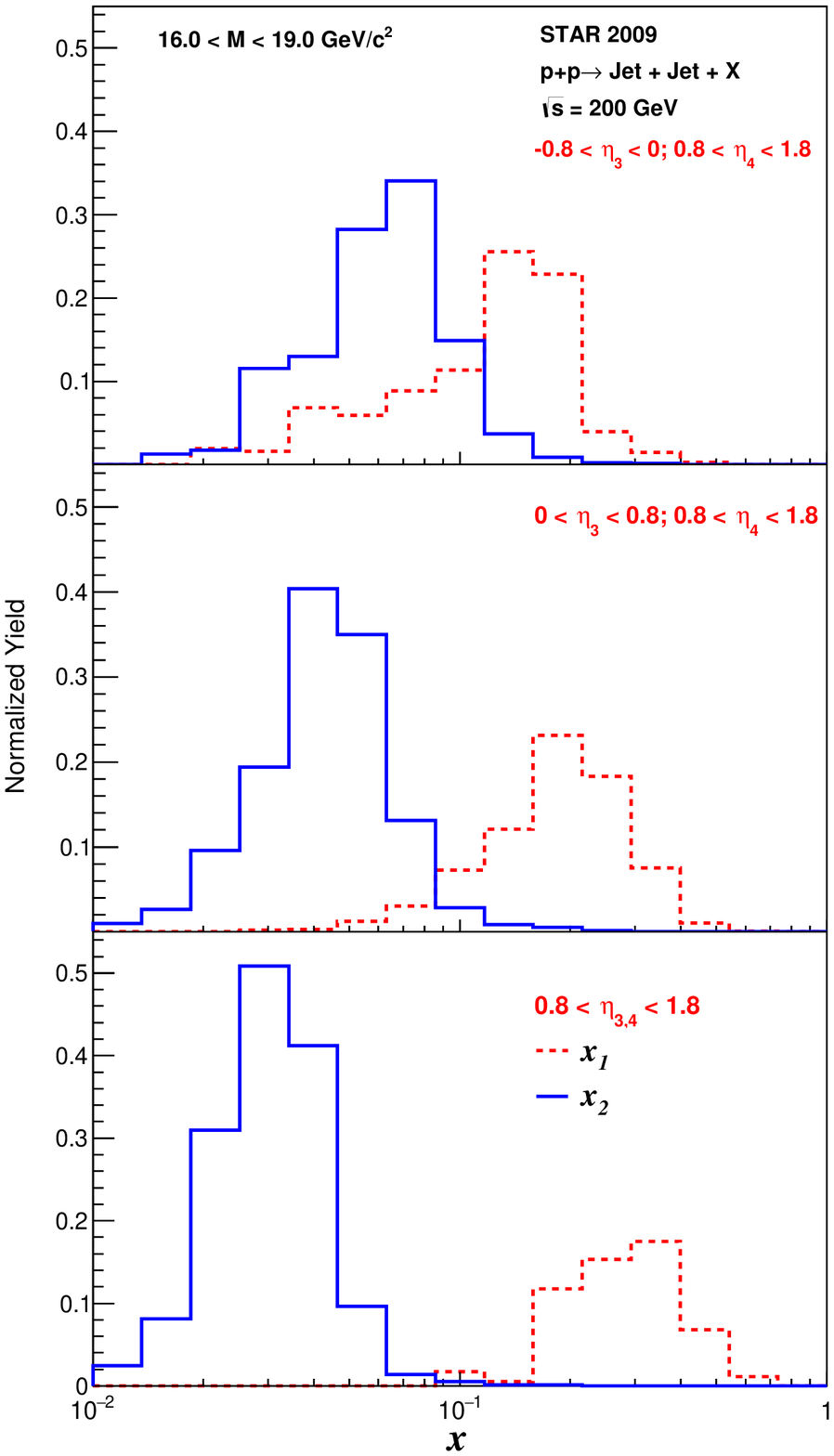}
  \caption{The distributions of the parton $x_1$ and $x_2$, which has been weighted by the partonic $\hat{a}_{LL}$, from \textsc{Pythia} detector level simulations at $\sqrt{s}$ = 200 GeV for different jet pseudorapidity ranges.}
  \label{fig:Dijet_Kinematics}
\end{minipage}%
\hfill
\begin{minipage}{.49\textwidth}
    \includegraphics[width=1\columnwidth]{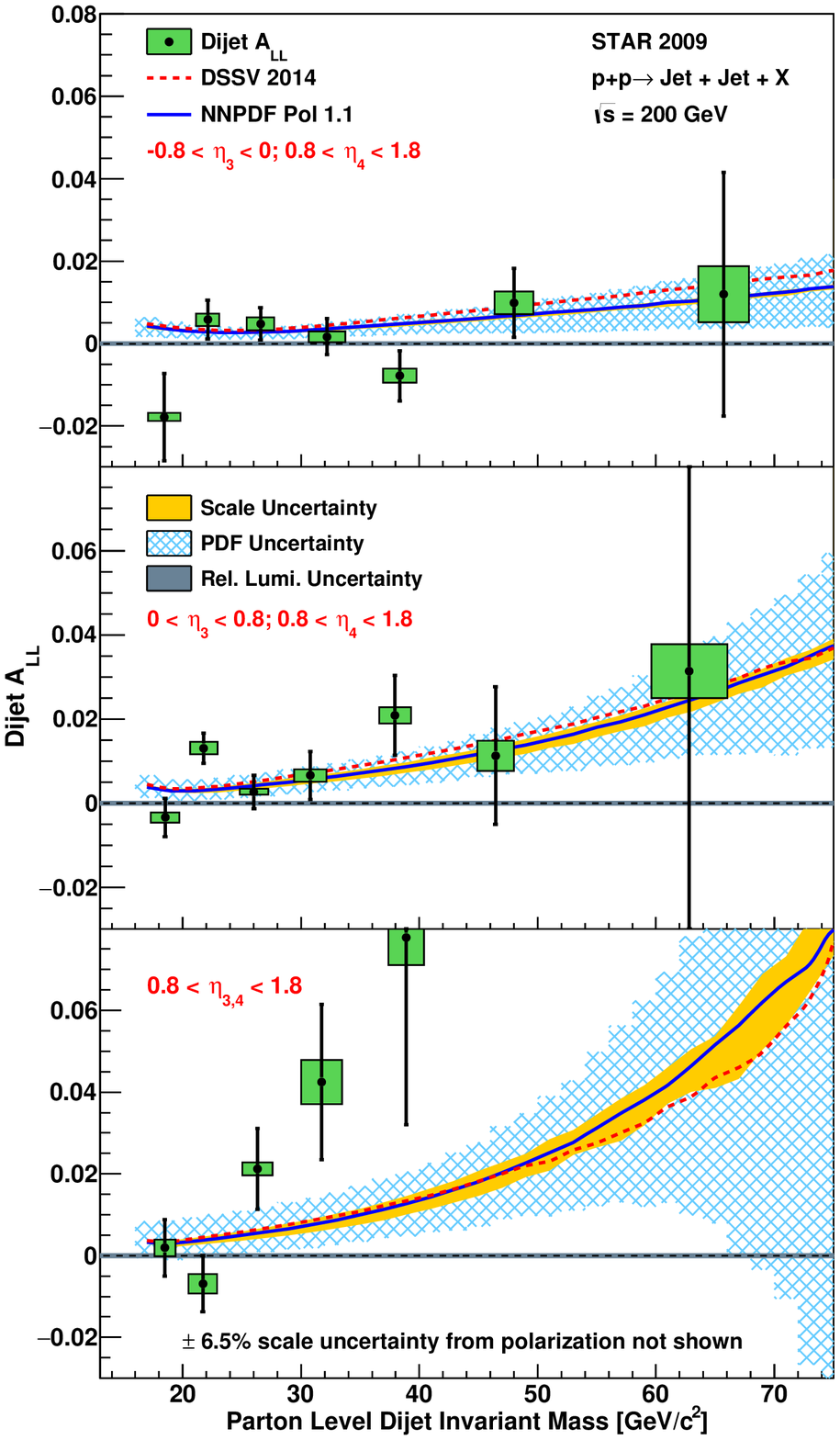}
    \caption{$A_{LL}$ as a function of parton-level invariant mass for dijets with the East Barrel-Endcap (top), West Barrel-Endcap (middle) and Endcap-Endcap (bottom) event topologies. The curves and uncertainty symbols are explained in the text.}
    \label{fig:ALL_DiffTopo}
\end{minipage}
\end{figure*}

The various event topologies probe different ranges of the momentum fractions, $x_1$ and $x_2$, carried by the partons that participate in the hard scattering, where $x_1$ is associated with the beam heading towards the EEMC. The distributions of $x_1$ and $x_2$ obtained from simulation for the three topologies discussed above are shown in Fig.~\ref{fig:Dijet_Kinematics}. The distributions are weighted by the partonic $\hat{a}_{LL}$~\cite{Craigie:1984tk} appropriate for each subprocess in order to indicate the regions of sensitivity to gluon polarization. They correspond to a sample of dijets from \textsc{Pythia} with detector-level invariant masses in the range  $16.0 < M < 19.0$~GeV/$c^2$, which is sensitive to the lowest momentum fraction values probed by this analyses.\par

The asymmetric nature of the collisions can be seen in the separation of the high- and low-$x$ distributions. They also extend to lower $x$ values than was possible with the mid-rapidity analysis. As expected, the separation in $x$ between the two distributions increases as the sum $\eta_3 + \eta_4$ increases, signaling the larger momentum asymmetry of the colliding partons. Compared to the analogous distributions generated for the STAR dijet measurements in the Barrel-Barrel topology under similar kinematic conditions, which provide sensitivity down to $x \sim 0.05$ \cite{Adamczyk:2016okk}, it is clear that extending the measurement into the Endcap region provides access to significantly lower values of $x$. Moreover, the large imbalance in the initial state momentum fractions, coupled with the shapes of well-established unpolarized PDF's, suggests that the low-$x$ peak is dominated by gluons, while the high-$x$ partons are most often valence quarks~\cite{Mukherjee:2012uz}.\par

Figure \ref{fig:ALL_DiffTopo} presents our values for $A_{LL}$ as a function of dijet mass, sorted by the same event topologies as were used in Tables \ref{table:Dijet_Systematic_Diff}. The $A_{LL}$ data shown have all been corrected back to the parton level, and are plotted at the mass-weighted average position of each dijet mass bin. The heights of the uncertainty boxes represent the total systematic uncertainty due to contributions from trigger and reconstruction bias, residual transverse polarization components in the beams, and uncertainties in the underlying events. The relative luminosity uncertainty is common to all points ({\it i.e.}, all asymmetries would move up or down by the same amount, independent of the asymmetry magnitude), and is represented by the small gray band on the horizontal axis. An overall vertical scale uncertainty of $6.5\%$, due to limitations in determining the absolute beam polarizations, is not shown. The widths of the uncertainty boxes represent the total systematic uncertainty associated with the corrected dijet invariant mass values and, in addition to contributions from the uncertainty on the individual jet corrections back to the parton level, include the uncertainties on calorimeter tower gains and efficiencies, as well as TPC momentum resolution and tracking efficiencies. A further uncertainty was added in quadrature to account for the differences among the \textsc{Pythia} tune sets. Underlying event effects, studied in both simulation and data, are included in the total systematic uncertainty.\par

Comparison of Figs.~\ref{fig:Dijet_Kinematics}--\ref{fig:ALL_DiffTopo} illustrates the advantages of studying correlation observables at forward pseudorapidity. Measurements using dijets constrain theoretical models over much narrower ranges of initial-state partonic momentum, compared to inclusive measurements, and thus provide more selective information on the shape ($x$-dependence) of helicity distributions. Sorting the events into different dijet topologies, based on the jet pseudorapidities, thereby enhances sensitivity of the data to selected regions in $x$, allowing cleaner sampling of the low-$x$ regions that are currently most poorly constrained in global analyses \cite{Nocera:2014gqa, deFlorian:2014yva}. Extending these measurements towards more forward rapidities increases the separation between $x_1$ and $x_2$, which not only probes even lower $x$ values, but also leads to a data sample dominated by the quark-gluon interactions of primary interest, that is, a high-$x$ (and therefore highly polarized) valence quark scattering from one of the abundant low-$x$ gluons.\par

\subsection{Comparison to theory}
The $A_{LL}$ asymmetry results presented in the figures are compared to two different theoretical model predictions. The theory curves were generated from the dijet production code of deFlorian \textit{et al.} \cite{deFlorian:1998qp}, using the DSSV2014 \cite{deFlorian:2014yva} and NNPDFpol1.1 \cite{Nocera:2014gqa} polarized PDF sets. The unpolarized PDF sets used to evaluate the denominator of the asymmetry calculations were MRST2008 \cite{Martin:2009iq} and NNPDF2.3 \cite{Ball:2013hta}, respectively. Uncertainty bands representing the sensitivity to factorization and renormalization scale (solid, yellow) and polarized PDF uncertainty (hatched, blue) were generated for the NNPDF results.\par 
The data are seen to be in generally good agreement with current theoretical model expectations, especially for the Barrel-Endcap events. Incorporating these results into the global analyses should lead to reduced uncertainties on the integrated value of $\Delta g(x)$, especially from contributions at smaller $x$.\par

\section{Summary}
In summary, first measurements of the longitudinal double-spin asymmetry $A_{LL}$ are presented for dijets detected at intermediate pseudorapidities. The dijets were recorded by the STAR collaboration in 2009, using polarized $pp$ collisions at $\sqrt{s}$ = 200~GeV. The final $A_{LL}$ results, corrected back to the parton-level and binned by dijet invariant mass for several pseudorapidity ranges, support the most recent DSSV and NNPDF predictions, both of which included the 2009 RHIC mid-rapidity inclusive jet and pion asymmetry data. The measurements reported here should provide new and tighter constraints on the magnitude, and especially the shape, of the gluon helicity distribution $\Delta g(x)$, particularly for $x < 0.05$, compared to previous studies.
With the increased statistics available from runs in 2012 and 2013 at $\sqrt{s}$ = 510~GeV, STAR data will help to further understand the behavior of $\Delta g(x)$ in the low $x$ region.\par

We thank the RHIC Operations Group and RCF at BNL, the NERSC Center at LBNL, and the Open Science Grid consortium for providing resources and support. This work was supported in part by the Office of Nuclear Physics within the U.S. DOE Office of Science, the U.S. National Science Foundation, the Ministry of Education and Science of the Russian Federation, National Natural Science Foundation of China, Chinese Academy of Science, the Ministry of Science and Technology of China and the Chinese Ministry of Education, the National Research Foundation of Korea, GA and MSMT of the Czech Republic, Department of Atomic Energy and Department of Science and Technology of the Government of India; the National Science Centre of Poland, National Research Foundation, the Ministry of Science, Education and Sports of the Republic of Croatia, RosAtom of Russia and German Bundesministerium fur Bildung, Wissenschaft, Forschung and Technologie (BMBF) and the Helmholtz Association.\par


\end{document}